\documentclass[11pt]{article}

\usepackage{bm}
\usepackage{url}
\usepackage{pdflscape}
\usepackage{amsmath}
\usepackage{amssymb}
\usepackage{caption}
\usepackage{subcaption}
\usepackage{booktabs}
\usepackage{tabularx}
\usepackage{arydshln}
\usepackage{xcolor}
\usepackage[hmargin=1in, vmargin=1in]{geometry}

\usepackage[symbol]{footmisc}

\usepackage{color}
\usepackage{graphicx}
\usepackage{multirow}
\usepackage{natbib}
\usepackage{rotating}

\usepackage{algpseudocode}  
\usepackage{algorithm} 
\usepackage{tcolorbox}  
\usepackage{arydshln}  

\usepackage[hidelinks]{hyperref}
\usepackage{orcidlink}

\usepackage{dsfont} 

\usepackage{mathtools}

\DeclarePairedDelimiterX{\expectarg}[1]{[}{]}{%
  \ifnum\currentgrouptype=16 \else\begingroup\fi
  \activatebar#1
  \ifnum\currentgrouptype=16 \else\endgroup\fi
}
\newcommand{\innermid}{\nonscript\;\delimsize\vert\nonscript\;}
\newcommand{\activatebar}{%
  \begingroup\lccode`\~=`\|
  \lowercase{\endgroup\let~}\innermid
  \mathcode`|=\string"8000
}

\newlength{\rowA}
\makeatletter

\begin{document}

\begin{center}

\textbf{\large The Long Shadow of Pandemic: \\ Understanding the lingering effects of cause-specific mortality shocks}
~\\  

\bigskip

Yanxin Liu$^{1}$\orcidlink{0000-0003-4244-9212} and Kenneth Q. Zhou$^{2}\footnote{Corresponding author. E-mail: \textit{kenneth.zhou@uwaterloo.ca}}$\orcidlink{0000-0002-6809-5208}
\end{center}

\begin{center}
$^1$Department of Finance, University of Nebraska-Lincoln, USA \\
$^2$Department of Statistics and Actuarial Science, University of Waterloo, Canada

\bigskip
October 14, 2026
\bigskip

\end{center}

\noindent\small\textbf{Abstract:} 
In the aftermath of the COVID-19 pandemic, empirical data have revealed that large-scale health crises not only cause immediate disruptions in mortality dynamics but also have persistent effects that may last for several years. Existing mortality models largely assume that mortality shocks are transitory and overlook how their effects can be long-lasting and heterogeneous across age groups and causes of death. In response to this limitation, we propose a novel stochastic mortality model that captures age- and cause-specific long-lasting effects of mortality jumps through a gamma-density-like decay function, estimated via a customized conditional maximum likelihood algorithm. Applying the model to recent U.S. mortality data, we reveal divergent persistence patterns across demographic groups and provide key insights into the tail risk profiles of life insurance and annuity products. Our scenario-based analyses further show that neglecting persistent shock effects can lead to suboptimal hedging, while the proposed model enables what-if testing to analyze such effects under potential future health crises.

\vspace{0.5cm}
\noindent\small\textbf{\textit{Keywords:}} \textit{Mortality shocks, Long-lasting pandemic effects, Stochastic mortality modeling, Cause-specific mortality, Natural hedging} \\

\section{Introduction}\label{sec:intro}

The COVID-19 pandemic has placed unprecedented financial strain on the insurance industry, with higher-than-expected claims due to increased mortality, particularly during pandemic peaks \citep{SwissRe2025}. Life insurers face additional uncertainty from volatile post-pandemic mortality patterns, long COVID's effects on future morbidity, and persistent behavioral and lifestyle changes, all of which complicate the assessment of long-term financial obligations \citep{RGA2021}. These challenges have forced insurers to reassess mortality assumptions and risk management strategies for handling future mortality shocks. Advanced mortality models that capture age-specific and cause-specific jump effects in mortality trends are therefore essential for improved risk assessment and portfolio management.

Modeling cause-specific mortality dynamics has drawn considerable interest in industry, especially in the wake of COVID-19. Recent technical reports, such as \cite{paglino2024soa} and \cite{ronora2025soa}, show that the pandemic’s impact on mortality is highly heterogeneous across causes of death and age groups. For older populations, COVID‑19 deaths account for the majority of excess mortality, with non‑COVID causes playing a comparatively smaller role. By contrast, younger individuals’ non‑COVID excess deaths, such as drug overdoses and vehicle accidents, constitute a substantial share of total excess mortality. These disparities highlight the importance of incorporating age‑specific and cause‑specific mortality dynamics in models intended to capture pandemic‑related mortality shocks.

The COVID-19 pandemic further reveals the complex interplay of age-specific disparities, cause-specific patterns, and long-lasting/lingering effects of mortality shocks. For instance, respiratory diseases generally exhibited lower-than-expected mortality, cardiovascular diseases showed persistent excess mortality, and external causes such as accidents and substance use remained elevated throughout the pandemic period \citep{paglino2024soa,IAA2021}. Moreover, recent data indicate that pandemic-related mortality effects extend beyond the acute crisis phase, driven by factors such as long COVID complications, delayed or foregone medical care, and lasting behavioral and socioeconomic changes \citep{paglino2024soa,SwissRe2025}. These effects collectively generate intricate mortality dynamics that conventional models struggle to capture.

Several studies have explored age- and cause-specific effects to better understand mortality dynamics, with early contributions including \cite{arnold2013forecasting, alai2015modelling, arnold2015causes, boumezoued2018cause, alai2018mind}. Building on this foundation, \cite{li2019modeling} employ hierarchical Archimedean copulas to model cause-specific mortality. \cite{li2019forecast} introduce a forecast reconciliation approach for capturing cause-specific mortality effects across different age groups and populations, while \cite{arnold2022short} use cointegration analysis within a vector error-correction model to analyze cause-specific mortality trends. More recently, a range of additional features, including elimination effects and socioeconomic factors, have been studied by \cite{huynh2024joint, yiu2023cause, yiu2024cause, dong2025compositional, varga2025forecasting, tanaka2025interpretable, villegas2025modelling, graziani2025age}. This growing body of research underscores the importance of isolating the causes of death and age effects for improved mortality forecasting and risk assessment.

Prior to the COVID-19 pandemic, a number of stochastic mortality models with jump effects were developed to understand extreme mortality risk. In a continuous-time setting, \cite{BiffisEnrico2005Apfd} and \cite{HainautDonatien2008MmwL} employ jump-diffusion and L\'evy-process specifications to characterize mortality jump effects. In a discrete-time setting, some researchers consider permanent jumps in which extreme events affect mortality dynamics indefinitely \citep{Cox2006,CoxSamuelH.2010MrmA,arik2023impact}, while others emphasize short-term transitory effects \citep{ChenCox2009,LinYijia2013PMSW,LiuLi2015,ozen2020transitory}. Regarding the distribution of jump occurrence, \cite{ChenCox2009} assume independent Bernoulli distributions, whereas \cite{Cox2006} consider Poisson jump counts. For the severity of jump effects, several researchers use a normal distribution \citep{ChenCox2009,LiuLi2015}, while double-exponential jumps are used by \cite{DengYinglu2012LRMa} and \cite{ChenHua2014AFoM}, and extreme value theory is applied by \cite{CHEN2010150}.

Since the COVID-19 pandemic, extensive research has been conducted on mortality modeling with pandemic-related effects. \cite{zhou2022multi} propose a three-parameter-level Lee-Carter extension to simulate future mortality scenarios with COVID-like effects. \cite{chen2022modeling} introduce a threshold jump framework, while \cite{carannante2022covid} analyze the effect of accelerated mortality shocks on life insurance contracts. \cite{robben2022assessing} and \cite{schnurch2022impact} study how pandemic data alter the calibration and projection of stochastic mortality models, and \cite{robben2024catastrophe} extend this line of work to a multi-population setting. \cite{VANBERKUM2025144} extend the Li-Lee model with a third pandemic layer calibrated to weekly mortality data, while \cite{Goes2025} extend the Lee-Carter model by introducing Bayesian mortality models with ``vanishing'' jump effects, where pandemic shocks are represented by serially dependent jump components.

Motivated by the empirical evidence and recent findings, this paper develops a novel stochastic mortality model that simultaneously captures age-specific severity, cause-specific patterns, and long-lasting persistence in mortality shocks. Unlike most existing models that treat jumps as either purely transitory or permanent, our approach introduces a flexible decay structure that allows shock effects to evolve differently across causes of death. The model is designed to serve practical applications in life insurance risk management, including liability valuation, de-risking strategies, and scenario-based stress testing. By providing a flexible modeling framework with empirically grounded features, this research equips actuaries and risk managers with tools for robust mortality risk assessment and long-term liability management.

The methodological contribution of this paper is a three-way parallel factors model with cause-specific lingering jump effects. The model decomposes log mortality rates into common trends, cause-specific deviations, and a novel jump component that captures both age-specific severity and cause-specific persistence through a gamma-density-like decay function. Applied to U.S. male mortality data spanning 1968--2023, the model reveals divergent pandemic impacts, such as causes with high initial severity but rapid decay and causes with lower severity but prolonged persistence. We also develop a conditional maximum likelihood estimation approach that efficiently handles the model's complexity while preserving estimation accuracy. Robustness checks confirm that the jump component successfully isolates pandemic effects without distorting long-term mortality trends, and model comparisons demonstrate the necessity of incorporating both age-specific and cause-specific heterogeneity.

Beyond its methodological contribution, the model offers practical insights for life insurance risk management. We examine natural hedging of longevity risk by balancing life insurance and annuity products in portfolio construction. The analysis reveals that optimal hedging requires accurate modeling of both age-specific and cause-specific mortality dynamics. Models that aggregate across causes or ages produce misspecified hedge ratios, substantially increasing portfolio risk relative to the optimal hedge. Scenario analyses demonstrate the model's flexibility in evaluating diverse mortality futures, including permanent mortality improvements from medical breakthroughs, endemic regimes with frequent mild shocks, and catastrophic events concentrated in specific age ranges. These capabilities provide insurers with tools for robust long-term risk planning under mortality uncertainty.

The remainder of this paper is organized as follows. Section \ref{sec:datamotiv} describes the cause-specific mortality data and presents empirical evidence of heterogeneous and long-lasting pandemic effects across age groups and causes of death. Section \ref{sec:models} introduces the 3WPF-CLJ model with its cause-specific lingering jump structure. Section \ref{sec:MdlAnl} presents the estimation results and analyzes parameter estimates, robustness, and model comparisons. Section \ref{sec:Application} examines applications to natural hedging and what-if scenario analysis for life insurance risk management. Lastly, Section \ref{sec:conclusion} concludes.

\section{Data and Motivation}\label{sec:datamotiv}

In this section, we provide empirical evidence motivating the development of age- and cause-specific mortality models capable of capturing long-lasting pandemic effects. We begin by describing the mortality data used throughout this study. We then investigate the pandemic's impact on aggregate mortality trends and the heterogeneity across age groups and causes of death. Lastly, we analyze the persistence of pandemic mortality shocks over the post-pandemic years.

\subsection{Data Description}\label{sec:datamotiv_data}

This paper utilizes mortality data from the U.S. Centers for Disease Control and Prevention (CDC) WONDER system.\footnote{The CDC WONDER system: \url{https://wonder.cdc.gov}} The dataset includes male and female mortality rates disaggregated by age group, year, and cause of death (CoD), obtained from two primary sources. The \emph{Compressed Mortality Dataset} provides death counts and population estimates from 1968 to 2016, spanning three revisions of the International Classification of Diseases (ICD) coding system: ICD-8 (1968--1978), ICD-9 (1979--1998), and ICD-10 (1999--2016). The \emph{Underlying Cause of Death Dataset} extends coverage from 1999 to 2023 using ICD-10 codes, with bridged-race categories for 1999--2020 and single-race categories for 2018--2023.

We combine these sources by using the Compressed Mortality Dataset for 1968--1998 and the Underlying Cause of Death Dataset for 1999--2023, yielding a unified sample of 56 years (1968--2023) across 13 age groups (0--1, 1--4, 5--9, 10--14, 15--19, 20--24, 25--34, 35--44, 45--54, 55--64, 65--74, 75--84, and 85+). Following \cite{arnold2022short}, we classify deaths into six CoD groups, with the first five corresponding to major disease categories and the sixth capturing all remaining causes including COVID-19, as summarized in Table~\ref{tab:CoD_Grouping}. We let $D_{x,t,c}$ denote the number of deaths for age group $x$ in year $t$ due to cause $c$. This quantity is structured as a three-dimensional array with dimensions $13 \times 56 \times 6$. We let $E_{x,t}$ denote the population exposures for age group $x$ in year $t$. Note that $E_{x,t}$ remains constant across CoD groups for a given age-year combination, and therefore is structured as a two-dimensional array with dimensions $13 \times 56$. The mortality rate for a specific CoD group $c$ is then defined as $m_{x,t,c} = D_{x,t,c} / E_{x,t}$, which serves as the primary quantity of interest in the remainder of this paper.

\begin{table}[h]
\centering
\begin{tabular}{lccc}
\hline \hline
\textbf{Coding system} & \textbf{ICD-10 Coding} & \textbf{ICD-9 Coding} & \textbf{ICD-8 Coding} \\
\hline
\textbf{Period} & 1999--2023 & 1979--1998 & 1968--1978 \\ \hline \hline
CoD 1: Infectious Diseases & A00-B99 & 001-139 & 001-136 \\
CoD 2: Cancer & C00-D48 & 140-239 & 140-239 \\
CoD 3: Circulatory Diseases & I00-I99 & 390-437 & 390-458 \\
CoD 4: Respiratory Diseases & J00-J98 & 460-519 & 460-519 \\
CoD 5: External Causes & V00-Y89 & E800-E999 & E810-E999 \\  
CoD 6: Other Causes + COVID & \multicolumn{3}{c}{All other causes (including COVID-19)} \\
\hline \hline
\end{tabular}
\caption{Summary of cause of death groupings across ICD coding systems.}
\label{tab:CoD_Grouping}
\end{table}

\subsection{Mortality Trends}
\label{sec:datamotiv_aggtrends}

\begin{figure}[th!]
    \centering
    \includegraphics[width=\linewidth]{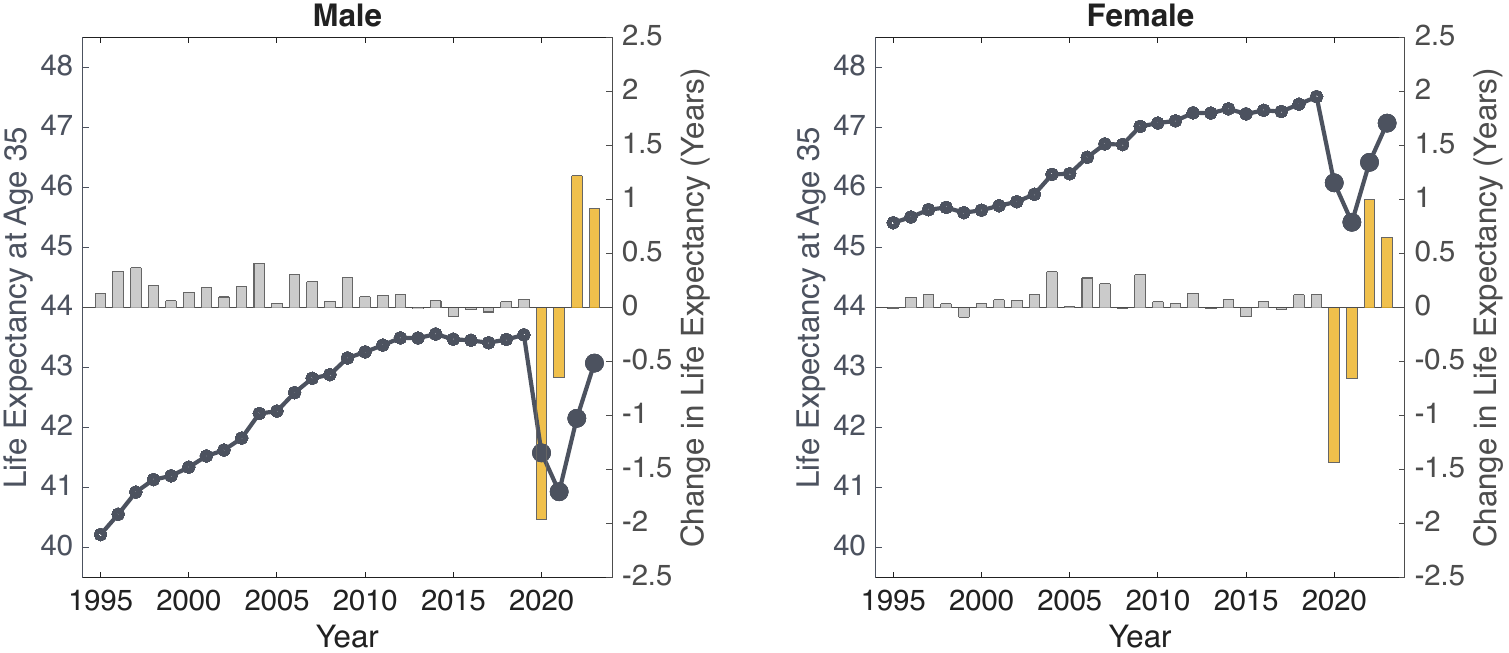}
    \caption{Period life expectancy at age 35 (dotted line) and year-over-year changes (vertical bars) for U.S. males and females, 1995--2023, with highlighted markers indicating the COVID-19 pandemic period.}
    \label{fig:LifeExp}
\end{figure}

Figure \ref{fig:LifeExp} presents the period life expectancy at age 35 from 1995 to 2023 for U.S. males and females, along with year-over-year changes in life expectancy shown on the right y-axis. To calculate these life expectancies, we apply spline interpolation to the log mortality rates $\ln(m_{x,t,c})$ along the age dimension for each cause and year, which preserves the age-specific patterns inherent to each cause and year. The period life expectancy for each year is then obtained by combining mortality across all causes of death.

Leading up to the COVID-19 pandemic, life expectancy exhibited a steady upward trend before 2010, followed by a period of stalled improvement up to 2019. Life expectancy at age 35 increased from approximately 40 years in 1995 to around 43.5 years by 2010 for males, and then remained largely flat until 2019. Females showed a similar pattern, with life expectancy rising from about 45.5 years in 1995 to approximately 47 years by 2010, and then plateauing through 2019. In 2020, a sharp decline disrupted this trajectory, marking the immediate and profound impact of the COVID-19 pandemic. The situation deteriorated further in 2021 as the pandemic continued. The year-over-year changes reveal that life expectancy decreased by approximately 2--2.5 years during the peak pandemic years for both males and females. In 2022 and 2023, mild recoveries in life expectancy were observed. Nevertheless, life expectancy at age 35 remained well below pre-pandemic levels, highlighting the pandemic's enduring effects on aggregated mortality trends.

\subsection{Heterogeneity by Age and Cause}
\label{sec:datamotiv_heterogeneity}

\begin{figure}[th!]
    \centering
    \includegraphics[width=\linewidth]{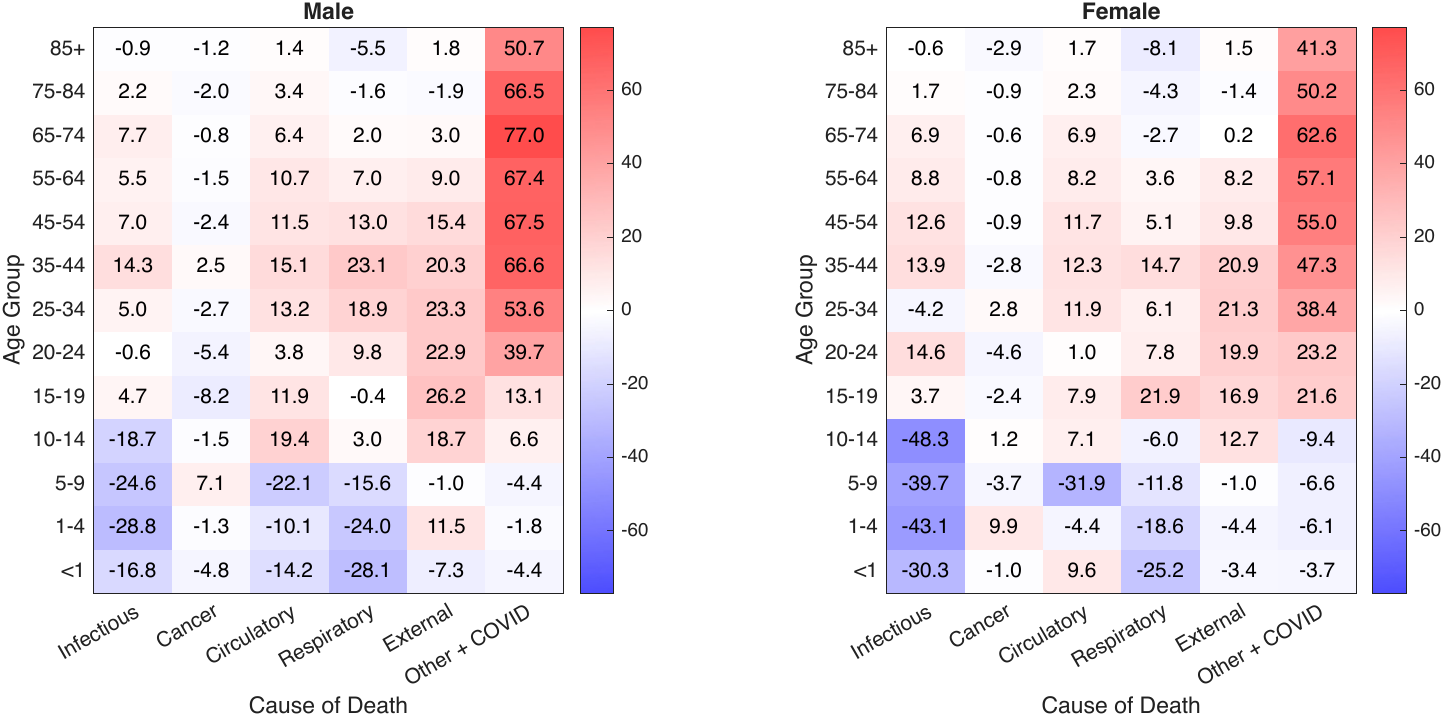}
    \caption{Percentage change in mortality rates from 2019 to 2020 by age group and cause of death for U.S. males (left) and females (right).}
    \label{fig:AgeCause_Heatmap}
\end{figure}

We now investigate how the pandemic's impact varied across age groups and causes of death. Figure \ref{fig:AgeCause_Heatmap} presents the percentage change in mortality rates between 2019 and 2020 across 13 age groups and 6 causes of death for both males and females. The color intensity indicates the magnitude and direction of change, with red representing increases in mortality and blue representing decreases. Numerical values displayed in each cell provide precise percentage changes for detailed comparison across age groups and causes.

The heatmap reveals substantial heterogeneity in the initial pandemic impact across age groups and causes. CoD 6 (Other + COVID) exhibited the most dramatic increases across nearly all age groups, with percentage changes ranging from approximately 40\% to 70\% for individuals aged 25 and older. This pattern is consistent across both genders and reflects the direct mortality burden of COVID-19. In contrast, CoD 1 (Infectious) showed clear decreases among younger age groups, with reductions reaching 20\% to 40\% for ages under 15, and young females experiencing stronger decreases than young males. CoD 5 (External) displayed notable increases concentrated in teenage and working-age groups (10--64 years), with changes ranging from 10\% to 25\%, and the impact was more pronounced for males than females.

For the remaining causes of death, the pandemic's initial impact was relatively modest. CoD 2 (Cancer) showed minimal changes across most age groups, with percentage changes typically within $\pm$5\%. CoD 3 (Circulatory) and CoD 4 (Respiratory) exhibited mixed patterns, with slight decreases among younger ages and modest increases among older ages. These heterogeneous patterns reveal that the pandemic represents more than just an additional cause of death but rather a systematic and differential shock to pre-existing mortality dynamics. This complexity underscores the need for developing age- and cause-specific modeling frameworks capable of capturing such heterogeneous mortality shocks.

\subsection{Evidence of Long-Lasting Effects}
\label{sec:datamotiv_longlasting}

\begin{figure}[th!]
    \centering
    \includegraphics[width=0.9\linewidth]{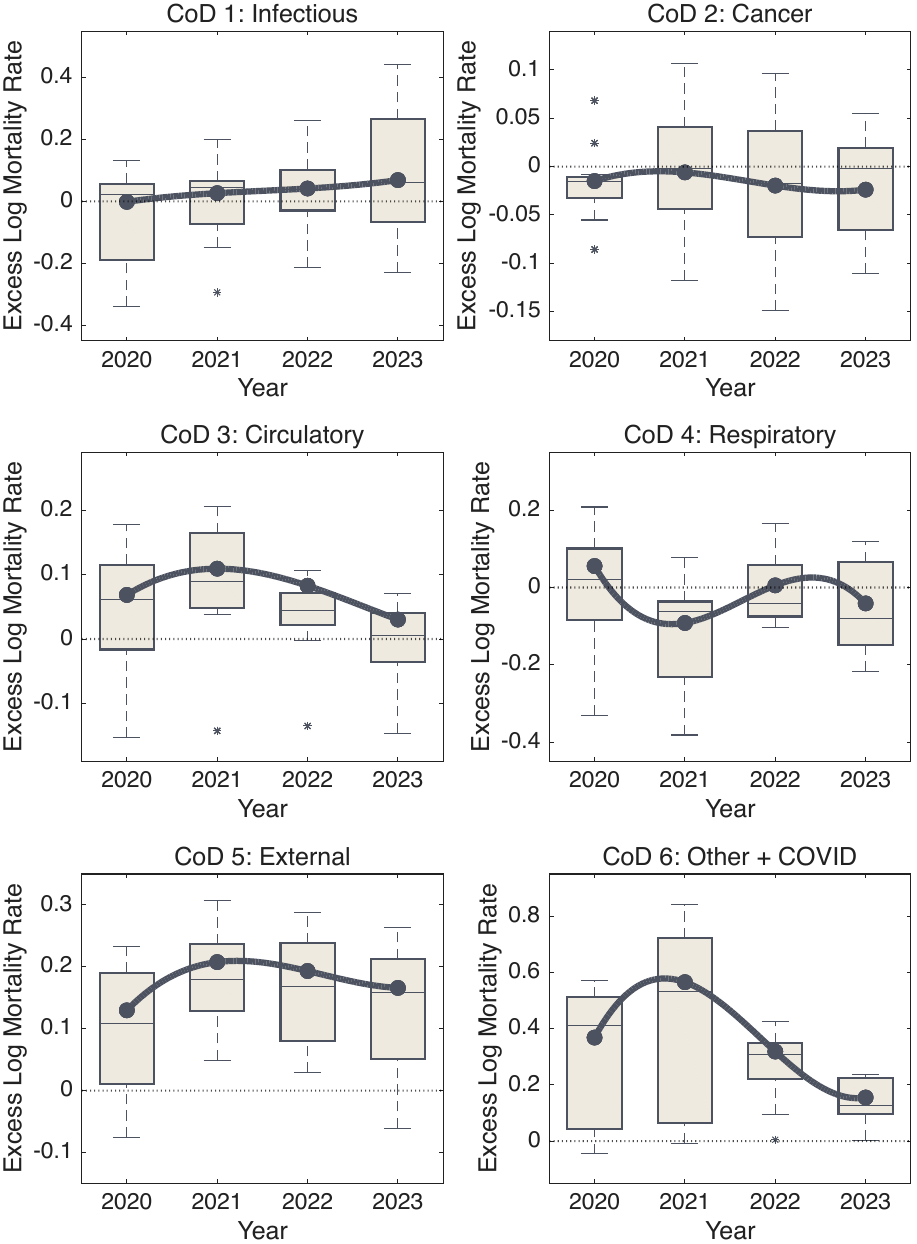}
    \caption{Excess log mortality rates in years 2020--2023 by cause of death for U.S. males; box plots show age distributions, dots show age-standardized means, and splines show temporal trajectories.}
    \label{fig:LongLast}
\end{figure}

Beyond the heterogeneous initial impact across age groups and causes, we also investigate whether the recent pandemic mortality shocks have long-lasting effects over multiple years, particularly focusing on the differences among causes of death. To quantify the long-lasting effects, we examine excess log mortality rates in the post-pandemic years, defined as
\begin{equation*} \label{eqn:excess_mortality}
    EM_{x,t,c} = \ln(m_{x,t,c}) - \ln(m_{x,t^{\ast},c}),
\end{equation*}
where $t^{\ast}$ is fixed at 2019 as the baseline year immediately prior to the pandemic, and $t$ ranges from 2020 to 2023. Figure \ref{fig:LongLast} presents the distribution of $EM_{x,t,c}$ for each cause during years 2020--2023. Each panel displays box plots of $EM_{x,t,c}$ across age groups, while the black dots indicate the age-standardized mean. A cubic spline curve is used to illustrate the trajectory of the age-standardized rates over time.

Figure \ref{fig:LongLast} reveals that persistence patterns differ substantially across the 6 causes of death. CoD 6 (Other + COVID) demonstrates the largest initial surge in 2020, peaking in 2021 and sharply declining afterward. CoD 3 (Circulatory) and CoD 5 (External) show more persistent effects, reaching their peak in 2021--2022 and experiencing slower decay afterward. CoD 1 (Infectious), CoD 2 (Cancer), and CoD 4 (Respiratory) display minimal fluctuations near zero throughout the period. These divergent patterns indicate that pandemic mortality shocks vary substantially in their persistence across causes of death.

\section{The Proposed Model}\label{sec:models}

This section introduces our three-way parallel factors model with cause-specific long-lasting/lingering mortality jump effects (3WPF-CLJ model). Motivated by the empirical findings from Section \ref{sec:datamotiv}, we develop a stochastic mortality model that captures age- and cause-specific jump effects with heterogeneous persistence patterns. We first present the model structure and distributional assumptions, and then derive an estimation approach tailored to the model along with the likelihood function based on mortality improvement rates.

\subsection{Model Structure} \label{sec:models_struc}

The proposed 3WPF-CLJ model decomposes the log mortality rate as follows:
\begin{equation}
\label{modelspec_deathrate}
\ln(m_{x,t,c}) = a_{x,c}+B_xK_t+\varphi_cb_xk_t+N_tJ_{x,c}^{(\tau)}+e_{x,t,c}.
\end{equation}
In this decomposition, the first component $a_{x,c}$ represents the level of mortality rate at age $x$ for cause of death (CoD) $c$. The second component is the product of $B_x$ and $K_t$, which represents the common factor shared by all causes of death. In particular, $K_t$ captures the variation of log mortality rates over time, and $B_x$ measures the sensitivity of log mortality rates to changes in $K_t$. The third component is the product of $\varphi_c$, $b_x$, and $k_t$, which reflects the long-term cause-specific deviation from the general trend in the cause of death, age, and time dimensions, respectively. The fourth component $N_tJ_{x,c}^{(\tau)}$ captures the jump effects, which we discuss in detail below. Finally, the fifth component $e_{x,t,c}$ is the error term. {We note that the six causes of death are not modeled independently, since they share the period factors and the jump indicator, and all-cause mortality is obtained as $\sum_c m_{x,t,c}$, which is coherent with the cause-specific forecasts by construction.}

The major novelty of the proposed model is the fourth component $N_tJ_{x,c}^{(\tau)}$, which captures the heterogeneous jump effects. We let $N_t=\mathds{1}_{\{ t \geq T_J\}}$ be an indicator variable, where $T_J$ is the year of pandemic occurrence, with $N_t=1$ {indicates} that a jump has occurred on or before year $t$, and $N_t=0$, otherwise. We let $J_{x,c}^{(\tau)}$ be the lingering jump variable at age $x$ attributed to cause of death $c$. The superscript $\tau$ denotes the time elapsed since the jump occurrence, computed as $\tau=t-T_J$. We then define the cause-specific lingering jump effect by
\begin{equation} \label{eqn: def_Jxtc_tau}
    J_{x,c}^{(\tau)} = \left\{ 
    \begin{array}{cl}
    0 & , \quad \tau < 0 \quad \text{or equivalently,} \quad t<T_J \\
    J_{x,c} & , \quad \tau = 0 \quad \text{or equivalently,} \quad t = T_J \\
    J_{x,c} \times \pi_c(\tau) & , \quad \tau > 0 \quad \text{or equivalently,} \quad t > T_J
    \end{array}
    \right.
\end{equation}
where $J_{x,c}$ represents the size of the initial jump for age $x$ attributed to cause of death $c$, and $\pi_c(\tau)$ represents the long-lasting effect for cause of death $c$. We remark that $J_{x,c}^{(\tau)}$ can capture the following three scenarios:
\begin{enumerate}
    \item Prior to jump occurrence ($t < T_J$), there is no jump effect to capture.
    
    \item When the jump occurs ($t=T_J$), the initial impact is captured by $J_{x,c}$.

    \item After jump occurrence ($t > T_J$), the long-lasting effect is captured by $\pi_c(\tau)$, a parametric function of time elapsed $\tau$.
\end{enumerate}

To incorporate the cause-specific long-lasting effects illustrated in Figure \ref{fig:LongLast}, we define
\begin{equation} \label{eqn: def_LLE}
    \pi_c(\tau) = \gamma_c \times \beta_c^{\alpha_c} \times \tau^{\alpha_c-1} \times e^{-\beta_c\tau},
\end{equation}
for $\tau > 0$, with magnitude parameter $\gamma_c$, shape parameter $\alpha_c$, and rate parameter $\beta_c$ that are all cause-specific. The magnitude parameter $\gamma_c$ supplements $J_{x,c}$ to adjust the size of the lingering jump effect. The shape parameter $\alpha_c$ controls the timing of the peak effect, allowing different causes of death to have varying response patterns to the same extreme mortality event. The rate parameter $\beta_c$ determines how rapidly the jump effect decays. It is worth noting that the specification of $\pi_c(\tau)$ is identical to a gamma density with parameters $\alpha_c$ and $\beta_c$, scaled by the magnitude parameter $\gamma_c$. Equation \eqref{eqn: def_Jxtc_tau} can be compactly expressed as
\begin{equation} \label{eqn: def_Jxtc_tau compact}
    J_{x,c}^{(\tau)} = J_{x,c} \times \pi_c(\tau),
\end{equation}
where $\pi_c(\tau)=0$ when $\tau<0$, and $\pi_c(\tau)=1$ when $\tau=0$.

We conclude this subsection by highlighting several key features of the proposed 3WPF-CLJ model. First, following the spirit of the Li-Lee model developed by \cite{LiLee2005}, the 3WPF-CLJ model incorporates both a common factor and a cause-specific factor. This design allows us to capture the general mortality trend while preserving the distinct features associated with different causes of death. Second, inspired by the three-way Lee-Carter model developed by \cite{Russolillo2011} and the four-way structure considered in \cite{cardillo2023mortality}, our model uses a single compact component $\varphi_c b_x k_t$ to capture the cause-specific deviation from the general trend. Finally, the 3WPF-CLJ model extends the transitory jump models introduced in \cite{LiuLi2015} by replacing $J_{x,t}$ with $J_{x,c}^{(\tau)}$. This extension allows the model to account not only for age-specific mortality jumps but also for long-lasting cause-specific effects.

\subsection{Model Assumptions} \label{sec:models_assum}

We now provide the distributional assumptions and the parameter constraints underlying the proposed model.
\begin{itemize}
    \item The period effect for the general trend $K_t$ follows a random walk with drift $D$:
    \[
    K_t = D+K_{t-1}+\eta_t
    \]
    where $\eta_t \sim N\left(0,\sigma_{\eta}^2\right)$. This assumption is widely adopted in stochastic mortality modeling, including the original Lee-Carter model \citep{LeeCarter1992} and the Li-Lee model \citep{LiLee2005}.
    
    \item The period effect for the cause-specific trend $k_t$ also follows a random walk with drift $d$:
    \[
    k_t = d+k_{t-1}+\xi_t
    \]
    where $\xi_t \sim N\left(0,\sigma_{\xi}^2\right)$. For simplicity, we assume $\eta_t$ and $\xi_t$ are uncorrelated. 
    
    \item The occurrence of a jump in each year follows an independent Bernoulli distribution; that is, $N_t \sim \text{Bern}(p)$, where $p$ denotes the probability of a jump occurring in a given year. We assume at most one jump occurs within the sample period.\footnote{This assumption is appropriate for our dataset, which spans 1968--2023 and contains only one major extreme mortality event (COVID-19). It allows us to focus on characterizing the long-lasting effects of this pandemic.} Under this assumption, the jump occurrence time $T_J$ follows a geometric distribution with probability mass function
    \[
    P(T_J = t_J) = P(N_1=\cdots=N_{t_J-1}=0,N_{t_J}=\cdots=N_{T}=1) = (1-p)^{t_J-1}\times p.
    \] 
    When no jump occurs, we have
    \[
    P(T_J \neq 1,\ldots, T) = P(N_1=\cdots=N_{T}=0)= (1-p)^{T}.
    \]
    
    \item The severity of the initial jump effect $J_{x,c}$ follows a normal distribution with mean varying across ages and causes of death. Due to the limited number of realized jumps in our dataset, we assume constant variance across all age-cause combinations. We denote the mean and variance by 
    \[
    \mu_{x,c}:=\text{E}\left(J_{x,c}\right)   \quad \text{and} \quad \sigma_{J}^2 := \text{Var}\left(J_{x,c}\right).
    \]
    Under this specification, the expectation and variance of the lingering jump effect $J_{x,c}^{(\tau)}$ are given by
    \begin{equation}
        \text{E}\left(J_{x,c}^{(\tau)}\right)= \mu_{x,c} \times \pi_c(\tau) 
    \end{equation}
    and
    \begin{equation}
    \text{Var}\left(J_{x,c}^{(\tau)}\right) = \sigma_{J}^2 \times (\pi_c(\tau))^2,
    \end{equation}
    respectively.
    
    \item The error term $e_{x,t,c}$ follows a normal distribution with zero mean and constant variance; that is, $e_{x,t,c}\sim N(0,\sigma_e^2)$. 
    \item { The age effects of the general trend $B_x$ and the cause-specific deviation $b_x$ satisfy the following parameter constraints:
    \[
    \sum\limits_xB_x = 1, \quad \text{ and } \quad  \sum\limits_xb_x=1.  
    \]
    Furthermore, we assume $\varphi_1 = 1$ so that the cause-specific deviation reflects the relative sensitivity to CoD 1.
    }
\end{itemize}

\subsection{Model Estimation} \label{sec:models_estim}

We conduct parameter estimation using the Route II estimation method, which has been shown to perform well for Lee-Carter type models \citep{HabermanRenshaw2012} and has been implemented in mortality models with transitory jump effects \citep{LiuLi2015}. Under the assumptions outlined in Section \ref{sec:models_assum}, the Route II approach can effectively reduce the number of parameters needed for estimation.

\subsubsection{Defining Mortality Improvement Rates}
To implement the Route II estimation method, we first introduce the notation\footnote{For ease of reference, Table~\ref{tab:notation} in Appendix~\ref{app:notation} summarizes the notation used throughout this paper.} 
\[
Z_{x,t,c} := \ln(m_{x,t,c})-\ln(m_{x,t-1,c})
\]
to represent the log mortality improvement rate at age $x$ in year $t$ for cause of death $c$. Taking the first difference of equation \eqref{modelspec_deathrate}, we obtain
\[
Z_{x,t,c} = B_x(K_t-K_{t-1})+\varphi_cb_x(k_t-k_{t-1})+N_tJ_{x,c}^{(\tau)}-N_{t-1}J_{x,c}^{(\tau-1)}+e_{x,t,c}-e_{x,t-1,c},
\]
which, under the random walk assumptions on $K_t$ and $k_t$, simplifies to    
\begin{equation}
\label{eqn:zxti simp}
Z_{x,t,c} = B_x(D+\eta_t)+\varphi_cb_x(d+\xi_t)+N_tJ_{x,c}^{(\tau)}-N_{t-1}J_{x,c}^{(\tau-1)}+e_{x,t,c}-e_{x,t-1,c}.
\end{equation}
Conditional on $N_{t-1}$ and $N_t$, $Z_{x,t,c}$ follows a normal distribution with mean equal to  
\begin{equation}
\label{eqn:zxti exp}
\begin{array}{cll}
\text{E}(Z_{x,t,c}|N_{t-1},N_t) &= B_xD+\varphi_cb_xd+\left(N_t\pi_c(\tau)-N_{t-1}\pi_c(\tau-1)\right)\mu_{x,c}   \\
 & = \left\{ 
    \begin{array}{ll}
    B_xD+\varphi_cb_xd & , \quad N_{t-1}=N_{t}=0 \\
    B_xD+\varphi_cb_xd+\mu_{x,c} & , \quad N_{t-1}=0, N_{t}=1 \\
    B_xD+\varphi_cb_xd + (\pi_c(\tau)-\pi_c(\tau-1))\times\mu_{x,c} & , \quad N_{t-1}=N_{t}=1
    \end{array}
    \right.
\end{array}
\end{equation}    
and variance equal to
\begin{equation}
\label{eqn:zxti var}
\begin{array}{cll}
\text{Var}(Z_{x,t,c}|N_{t-1},N_t) &= B_x^2\sigma_{\eta}^2 + \varphi_c^2b_x^2\sigma_{\xi}^2 + \left(N_t\pi_c(\tau)-N_{t-1}\pi_c(\tau-1)\right)^2\sigma_{J}^2 + 2\sigma_{e}^2  \\
 & = \left\{ 
    \begin{array}{ll}
    B_x^2\sigma_{\eta}^2 + \varphi_c^2b_x^2\sigma_{\xi}^2  + 2\sigma_{e}^2 & , \quad N_{t-1}=N_{t}=0 \\
    B_x^2\sigma_{\eta}^2 + \varphi_c^2b_x^2\sigma_{\xi}^2 + \sigma_{J}^2 + 2\sigma_{e}^2 & , \quad N_{t-1}=0, N_{t}=1 \\
    B_x^2\sigma_{\eta}^2 + \varphi_c^2b_x^2\sigma_{\xi}^2 + \left(\pi_c(\tau)-\pi_c(\tau-1)\right)^2\sigma_{J}^2 + 2\sigma_{e}^2 & , \quad N_{t-1}=N_{t}=1.
    \end{array}
    \right.
\end{array}
\end{equation}    

Depending on the values of $N_{t-1}$ and $N_t$, the conditional distribution of $Z_{x,t,c}$ corresponds to one of three scenarios. The first scenario is $N_{t-1}=N_t=0$, which indicates that no jump has occurred and thus the jump effect can be excluded from the mean and variance. In the second scenario, we have $N_{t-1}=0$ and $N_t=1$, which indicate that a jump occurs in year $t$ and the jump effect enters the model but with no persistence yet. In the last scenario, $N_{t-1}=N_t=1$ indicates that a jump occurred prior to year $t$ and the long-lasting effect of the jump needs to be incorporated into the mean and variance.

\subsubsection{Designing a Matrix Representation}
The expressions in equations \eqref{eqn:zxti simp}, \eqref{eqn:zxti exp}, and \eqref{eqn:zxti var} can be expressed compactly using a matrix representation. Because of the three-dimensional structure of our dataset, the log mortality improvement rates are indexed by age, time, and cause. We suppress these three dimensions into a matrix representation as follows. 

\par\smallskip
\noindent{\bf \underline{Suppressing the Age Dimension}}\\
We first suppress the age dimension by using the notation 
\[
     \vec{Z}_{t,c} =
    \left(\begin{matrix}
Z_{1,t,c} \\ 
Z_{2,t,c} \\
\vdots \\
Z_{X,t,c}
\end{matrix} \right), \quad
     \vec{J}_{c} =
    \left(\begin{matrix}
J_{1,c} \\ 
J_{2,c} \\
\vdots \\
J_{X,c}
\end{matrix} \right), \quad
     \vec{\mu}_{c} =
    \left(\begin{matrix}
\mu_{1,c} \\ 
\mu_{2,c} \\
\vdots \\
\mu_{X,c}
\end{matrix} \right), \quad
     \vec{e}_{t,c} =
    \left(\begin{matrix}
e_{1,t,c} \\ 
e_{2,t,c} \\
\vdots \\
e_{X,t,c}
\end{matrix} \right)
\]
to represent vectors of log mortality improvement rates, jump effects, expected jump effects, and errors in year $t$ for the cause of death $c$, and
\[
\vec{B} = (B_1,B_2,\ldots,B_X)^\top, \quad
\vec{b} = (b_1,b_2,\ldots,b_X)^\top, \quad
\vec{1}_X = (1,1,\ldots,1)^\top
\]
to represent vectors of $B_x$, $b_x$, and ones with $X$ elements, respectively. The vector of lingering jump effects in year $t$ for cause of death $c$ is then given by $\vec{J}_{c}^{(\tau)} = \vec{J}_{c}\times \pi_c(\tau)$. It follows that equation \eqref{eqn:zxti simp} can be expressed as
\[    
\vec{Z}_{t,c} = \vec{B}(D+\eta_t)+\varphi_c\vec{b}(d+\xi_t)+N_t\vec{J}_{c}^{(\tau)}-N_{t-1}\vec{J}_{c}^{(\tau-1)}+\vec{e}_{t,c}-\vec{e}_{t-1,c}.
\]

\par\smallskip
\noindent{\bf \underline{Suppressing the Cause Dimension}}\\
Proceeding similarly, we suppress the cause dimension by using
\[
     \vec{Z}_{t} =
    \left(\begin{matrix}
\vec{Z}_{t,1} \\ 
\vec{Z}_{t,2} \\
\vdots \\
\vec{Z}_{t,C}
\end{matrix} \right), \quad
     \vec{J} =
    \left(\begin{matrix}
\vec{J}_{1} \\ 
\vec{J}_{2} \\
\vdots \\
\vec{J}_{C}
\end{matrix} \right), \quad
     \vec{\mu} =
    \left(\begin{matrix}
\vec{\mu}_{1} \\ 
\vec{\mu}_{2} \\
\vdots \\
\vec{\mu}_{C}
\end{matrix} \right), \quad
     \vec{e}_{t} =
    \left(\begin{matrix}
\vec{e}_{t,1} \\ 
\vec{e}_{t,2} \\
\vdots \\
\vec{e}_{t,C}
\end{matrix} \right)
\]
to represent vectors of $Z_{x,t,c}$, $J_{x,c}$, $\text{E}(J_{x,c})$, and $e_{x,t,c}$ in year $t$, and
\[
\vec{\varphi} = (\varphi_1,\varphi_2,\ldots,\varphi_C)^\top, \quad
\vec{\pi}(\tau) = (\pi_1(\tau), \pi_2(\tau),\ldots, \pi_C(\tau))^\top, \quad
\vec{1}_C = (1,1,\ldots,1)^\top
\]
to represent vectors of $\varphi_c$, $\pi_c(\tau)$, and ones with $C$ elements, respectively. The lingering jump effect in year $t$ can be expressed as 
\[
\vec{J}^{(\tau)} = \left( \vec{\pi}(\tau) \otimes \vec{1}_X \right) \circ \vec{J},
\]
where $\circ$ denotes element-wise multiplication. The vector of log mortality improvement rates in year $t$ then follows
\[   
\begin{array}{cl}
\vec{Z}_{t}  &  = \vec{1}_{C} \otimes \vec{B} \times (D+\eta_t)+\vec{\varphi}\otimes\vec{b} \times (d+\xi_t)+N_t\vec{J}^{(\tau)}-N_{t-1}\vec{J}^{(\tau-1)}+\vec{e}_{t}-\vec{e}_{t-1} \\[0.3cm]
     & = \vec{1}_{C} \otimes \vec{B} \times (D+\eta_t)+\vec{\varphi}\otimes\vec{b} \times (d+\xi_t) +\left(\left(N_t\vec{\pi}(\tau)-N_{t-1}\vec{\pi}(\tau-1)\right)\otimes \vec{1}_X\right)\circ\vec{J}+\vec{e}_{t}-\vec{e}_{t-1},
\end{array}
\]
which provides a compact matrix representation of equation \eqref{eqn:zxti simp}. Conditional on $N_{t-1}$ and $N_t$, $\vec{Z}_t$ follows a multivariate normal distribution with mean
\[
\text{E}\left(\vec{Z}_{t}|N_{t-1},N_t\right)  = \vec{1}_{C} \otimes \vec{B}\times D+\vec{\varphi}\otimes\vec{b}\times d+\left(\left(N_t\vec{\pi}(\tau)-N_{t-1}\vec{\pi}(\tau-1)\right)\otimes \vec{1}_X\right)\circ\vec{\mu}
\]
and variance-covariance matrix
\[
\begin{array}{rcl}
\text{Var}\left(\vec{Z}_{t}|N_{t-1},N_t\right)   & = &  \left(\vec{1}_{C} \otimes \vec{B}\right)\left(\vec{1}_{C} \otimes \vec{B}\right)^\top \times\sigma_{\eta}^2  + \left(\vec{\varphi}\otimes\vec{b}\right)\left(\vec{\varphi}\otimes\vec{b}\right)^\top \times\sigma_{\xi}^2 \\[0.2cm]
& & + {\bf L}_t \Sigma_J {\bf L}_t^\top  + 2\times\mathbf{I}_{XC}\times \sigma_e^2,
\end{array}
\]
where $\Sigma_J = {\bf I}_{XC} \times \sigma_J^2$ by definition, and ${\bf L}_t$ is an $(XC)$-by-$(XC)$ diagonal matrix with diagonal elements $\left(N_t\vec{\pi}(\tau)-N_{t-1}\vec{\pi}(\tau-1)\right)\otimes \vec{1}_X$ and zeros elsewhere.

\par\smallskip
\noindent{\bf \underline{Suppressing the Time Dimension}}\\
Finally, we specify the full vector of log mortality improvement rates, $\vec{Z}$, by suppressing the time dimension. We define 
\[
    \vec{Z} =
    \left(\begin{matrix}
\vec{Z}_{2} \\ 
\vec{Z}_{3} \\
\vdots \\
\vec{Z}_{T}
\end{matrix} \right), \quad
    \vec{\eta} =
    \left(\begin{matrix}
\eta_{2} \\ 
\eta_{3} \\
\vdots \\
\eta_{T}
\end{matrix} \right), \quad
    \vec{\xi} =
    \left(\begin{matrix}
\xi_{2} \\ 
\xi_{3} \\
\vdots \\
\xi_{T}
\end{matrix} \right), \quad
    \vec{N} =
    \left(\begin{matrix}
N_{1} \\ 
N_{2} \\
\vdots \\
N_{T}
\end{matrix} \right), \quad
    \vec{\pi} =
    \left(\begin{matrix}
\vec{\pi}(1-t_J) \\ 
\vec{\pi}(2-t_J) \\ 
\vdots \\
\vec{\pi}(T-t_J) \\ 
\end{matrix} \right), \quad
    \vec{e} =
    \left(\begin{matrix}
\vec{e}_{1} \\ 
\vec{e}_{2} \\
\vdots \\
\vec{e}_{T}
\end{matrix} \right).
\]
The full vector of log mortality improvement rates can be expressed as
\[
\begin{array}{rl}
\displaystyle \vec{Z} = & \displaystyle (D\times \vec{1}_{T-1}+\vec{\eta}) \otimes \vec{1}_{C} \otimes \vec{B} + (d\times \vec{1}_{T-1}+\vec{\xi}) \otimes \vec{\varphi}\otimes\vec{b} \\[0.2cm]
  & \displaystyle + \left( \left({\bf I}^\ast_C \left( (\vec{N}\otimes\vec{1}_C)\circ \vec{\pi} \right) \right) \otimes \vec{1}_X \right) \circ \left( \vec{1}_{T-1}\otimes \vec{J}\right)  + {\bf I}^\ast_{XC}\vec{e},
\end{array}
\]
where 
\[
{\bf I}^\ast_C = \left( \begin{matrix}
    -{\bf I}_{C} & {\bf I}_{C} & {\bf 0} & \cdots  & {\bf 0} & {\bf 0} \\
    {\bf 0} & -{\bf I}_{C} &  {\bf I}_{C} & \cdots & {\bf 0} & {\bf 0} \\
    \vdots & \vdots & \ddots & \ddots & \vdots & \vdots  \\
    {\bf 0} & {\bf 0} & \cdots& \cdots  & -{\bf I}_{C} &  {\bf I}_{C} \\
\end{matrix} \right)
\]
is a sparse block matrix with $(T-1)$ rows and $T$ columns of blocks, where each block ${\bf I}_{C}$ is a $C$-by-$C$ identity matrix. Similarly, ${\bf I}^\ast_{XC}$ is a sparse block matrix with blocks ${\bf I}_{XC}$, where ${\bf I}_{XC}$ is an $XC$-by-$XC$ identity matrix. 

Conditional on $\vec{N}$, one can show that the expectation and variance of $\vec{Z}$ are given by
\begin{equation} \label{eqn: cond_exp_full_z}    
\text{E}(\vec{Z}|\vec{N}) = \displaystyle (D\times \vec{1}_{T-1}) \otimes \vec{1}_{C} \otimes \vec{B} + (d\times \vec{1}_{T-1}) \otimes \vec{\varphi}\otimes\vec{b} + \left( \left({\bf I}^\ast_C \left( (\vec{N}\otimes\vec{1}_C)\circ \vec{\pi} \right) \right) \otimes \vec{1}_X \right) \circ \left( \vec{1}_{T-1}\otimes \vec{\mu}\right)  
\end{equation}
and
\begin{equation} \label{eqn: cond_var_full_z}   
\begin{array}{rl}
\displaystyle \text{Var}(\vec{Z}|\vec{N}) = & \displaystyle \Sigma_\eta \otimes \vec{1}_C\vec{1}_C^\top \otimes \vec{B}\vec{B}^\top+\Sigma_\xi \otimes \vec{\varphi}\vec{\varphi}^\top \otimes \vec{b}\vec{b}^\top \\[0.2cm]
& \displaystyle + {\bf L} \left( (\vec{1}_{T-1}\vec{1}_{T-1}^\top) \otimes \Sigma_J \right) {\bf L}^\top + {\bf I}^\ast_{XC}\Sigma_e({\bf I}^\ast_{XC})^\top,
\end{array}
\end{equation}
respectively, where $\Sigma_\eta = {\bf I}_{T-1}\times \sigma_\eta^2$, $\Sigma_\xi = {\bf I}_{T-1}\times \sigma_\xi^2$, $\Sigma_e = {\bf I}_{T}\times \sigma_e^2$, and ${\bf L}$ is a diagonal matrix with diagonal elements ${\bf I}^\ast_C \left( (\vec{N}\otimes\vec{1}_C)\circ \vec{\pi} \right)$ and zeros elsewhere.

\subsubsection{Deriving the Log-Likelihood Function}
For notational convenience, we use $\vec{\theta}$ to represent the vector of all parameters in the proposed model, comprising parameters associated with the general mortality dynamic ($\vec{B}, D, \sigma_{\eta}$), cause-specific mortality dynamic ($\vec{b}, d, \sigma_{\xi}$), lingering jump effects ($\vec{\mu},\sigma_J,\vec{\alpha},\vec{\beta},\vec{\gamma},p$), and the random noise ($\sigma_e$). 

The log-likelihood function under the Route II approach is given by
\begin{equation} \label{eqn: loglik}    
l(\vec{\theta}) = \ln\left( f\left(\vec{z} \, ; \,\vec{\theta} \right) \right).
\end{equation}
By the law of total probability, the density function $f\left(\vec{z} \, ; \,\vec{\theta} \right)$ can be expressed as a sum over all jump occurrence patterns:
\begin{equation}    
\begin{array}{ccl}
\displaystyle f\left(\vec{z} \, ; \,\vec{\theta} \right) & =  & \sum\limits_{\vec{N}} f\left(\vec{z} \, | \, \vec{N} \,;\vec{\theta} \right)\times P(\vec{N}) \\[0.2cm]
\displaystyle     & = & f\left(\vec{z} \, | \, N_1=\cdots=N_T=0 \, ; \,\vec{\theta} \right)\times P(N_1=\cdots=N_T=0) \\[0.2cm]
\displaystyle     &  & + \, f\left(\vec{z} \, | \, N_1=\cdots=N_{T-1}=0, N_{T}=1 \, ; \,\vec{\theta} \right)\times P(N_1=\cdots=N_{T-1}=0, N_T=1) \\[0.2cm]
\displaystyle     &  & + \, \cdots \\[0.2cm]
\displaystyle     &  & + \, f\left(\vec{z} \, | \, N_1=\cdots=N_T=1 \, ; \,\vec{\theta} \right)\times P(N_1=\cdots=N_T=1).
\end{array} \label{eqn: complete joint density v1}
\end{equation}
Substituting the assumption on jump occurrence from Section \ref{sec:models_assum}, we get
\begin{equation}
    \begin{array}{cclc}
\displaystyle f\left(\vec{z} \, ; \,\vec{\theta} \right) & = & f\left(\vec{z} \, | \, N_1=\cdots=N_T=0 \, ; \,\vec{\theta} \right)\times (1-p)^T & \fbox{\text{No Jump}}\\[0.2cm]
\displaystyle     &  & + \, f\left(\vec{z} \, | \, N_1=\cdots=N_{T-1}=0, N_{T}=1 \, ; \,\vec{\theta} \right)\times (1-p)^{T-1}\times p & \fbox{\text{Jump in Year $T$}}\\[0.2cm]
\displaystyle     &  & + \, \cdots & \vdots\\[0.2cm]
\displaystyle     &  & + \, f\left(\vec{z} \, | \, N_1=\cdots=N_T=1 \, ; \,\vec{\theta} \right)\times p, & \fbox{\text{Jump in Year 1}}
\end{array} \label{eqn: complete joint density v2}
\end{equation}
Therefore, $f\left(\vec{z} \, ; \,\vec{\theta} \right)$ can be interpreted as a Gaussian mixture model comprising different jump occurrence scenarios. The mean and variance of each conditional distribution are given by equations \eqref{eqn: cond_exp_full_z} and \eqref{eqn: cond_var_full_z}, respectively.

\section{Model Analysis} \label{sec:MdlAnl}

This section presents comprehensive analyses of the proposed 3WPF-CLJ model. We first present the estimation results obtained using the Broyden-Fletcher-Goldfarb-Shanno (BFGS) algorithm, with implementation details provided in Appendix \ref{app:algorithm}. We then examine the model's robustness across different calibration windows, with particular focus on the influence of the data in COVID years. Finally, we conduct model comparisons to demonstrate the importance of incorporating long-lasting/lingering jump effects and the necessity of allowing cause-specific heterogeneity in jump effects.

\subsection{Estimation Results} \label{sec:MdlAnl_Est}

We fit the proposed 3WPF-CLJ model to { both U.S. male population and female population. We illustrate the results using} U.S. male mortality data from 1968 to 2023, as described in Section \ref{sec:datamotiv}\footnote{The estimation results for U.S. female population are provided in Appendix \ref{app: U.S. female result}.}. Recall that our dataset comprises 13 age groups, 6 causes (with COVID-19 deaths categorized in CoD 6), and 56 years of observations. {The maximized log-likelihood function is $4.5484 \times 10^3$, with 131 free parameters, yielding an AIC of $-8.8348 \times 10^3$ and a BIC of $-8.0011\times 10^3$.}

\begin{figure}
    \centering
    \includegraphics[width=\textwidth]{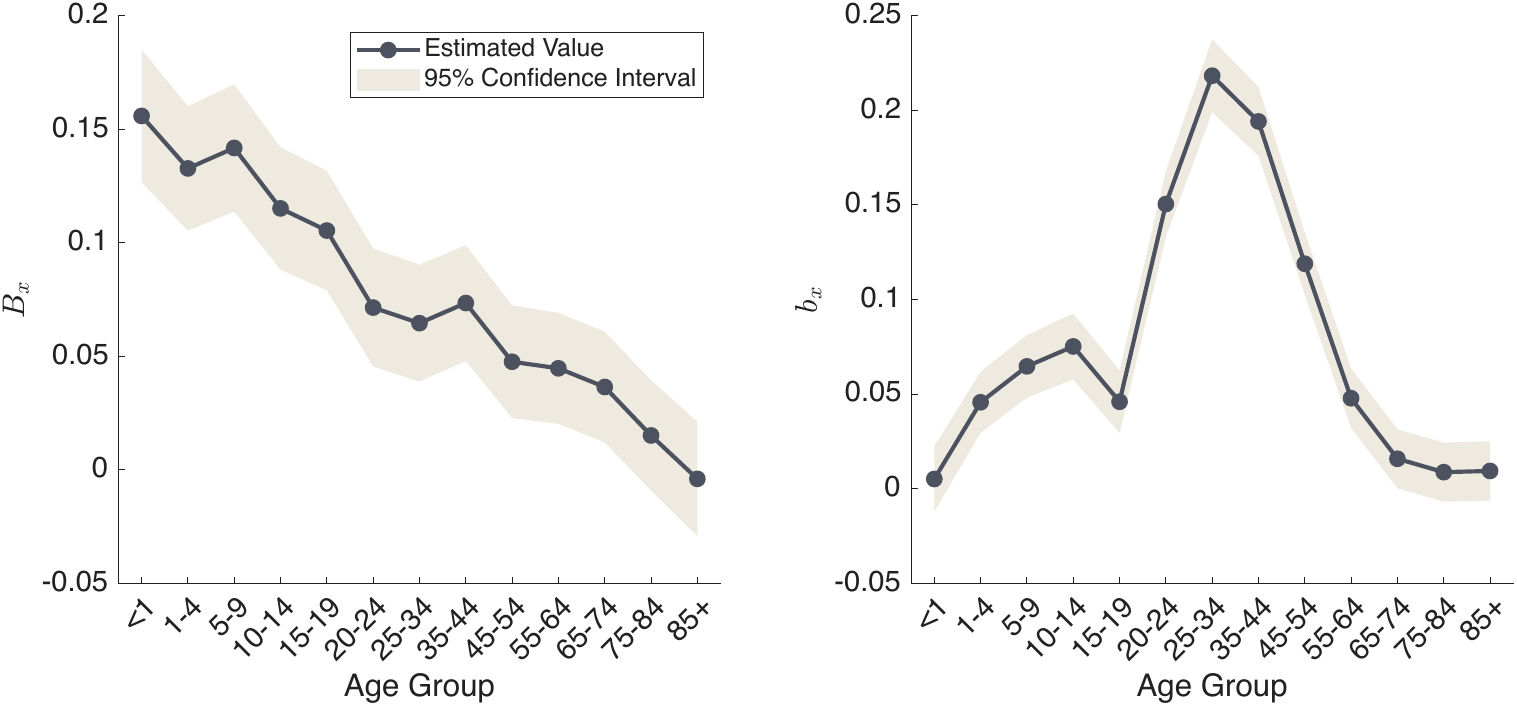}
    \caption{Estimated values of $B_x$ and $b_x$ with 95\% confidence intervals from the 3WPF-CLJ model fitted to U.S. male mortality data from 1968 to 2023.}
    \label{fig:Bxbx}
\end{figure}

Figure \ref{fig:Bxbx} presents the estimated values and 95\% confidence intervals for the common age pattern $B_x$ and the cause-specific age deviations $b_x$. The downward-sloping pattern of $B_x$ in the left panel of Figure \ref{fig:Bxbx} reveals a negative correlation between mortality improvement and age, indicating that older individuals experience smaller mortality improvements than younger individuals. In contrast, the right panel of Figure \ref{fig:Bxbx} shows a markedly different pattern for $b_x$, resembling a mixture of two normal densities with peaks at ages 10--14 and 25--34. {Since $\varphi_1$ for CoD 1 (Infectious), set to one for identification, is far greater than $\varphi_c$ for other CoDs (see Table \ref{tab:EstOthers}), the pattern of $b_x$ mainly captures deviations from CoD 1, which in our dataset is not related to the COVID-19 pandemic.}

\begin{figure}[ht!]
    \centering
    \includegraphics[width=\textwidth]{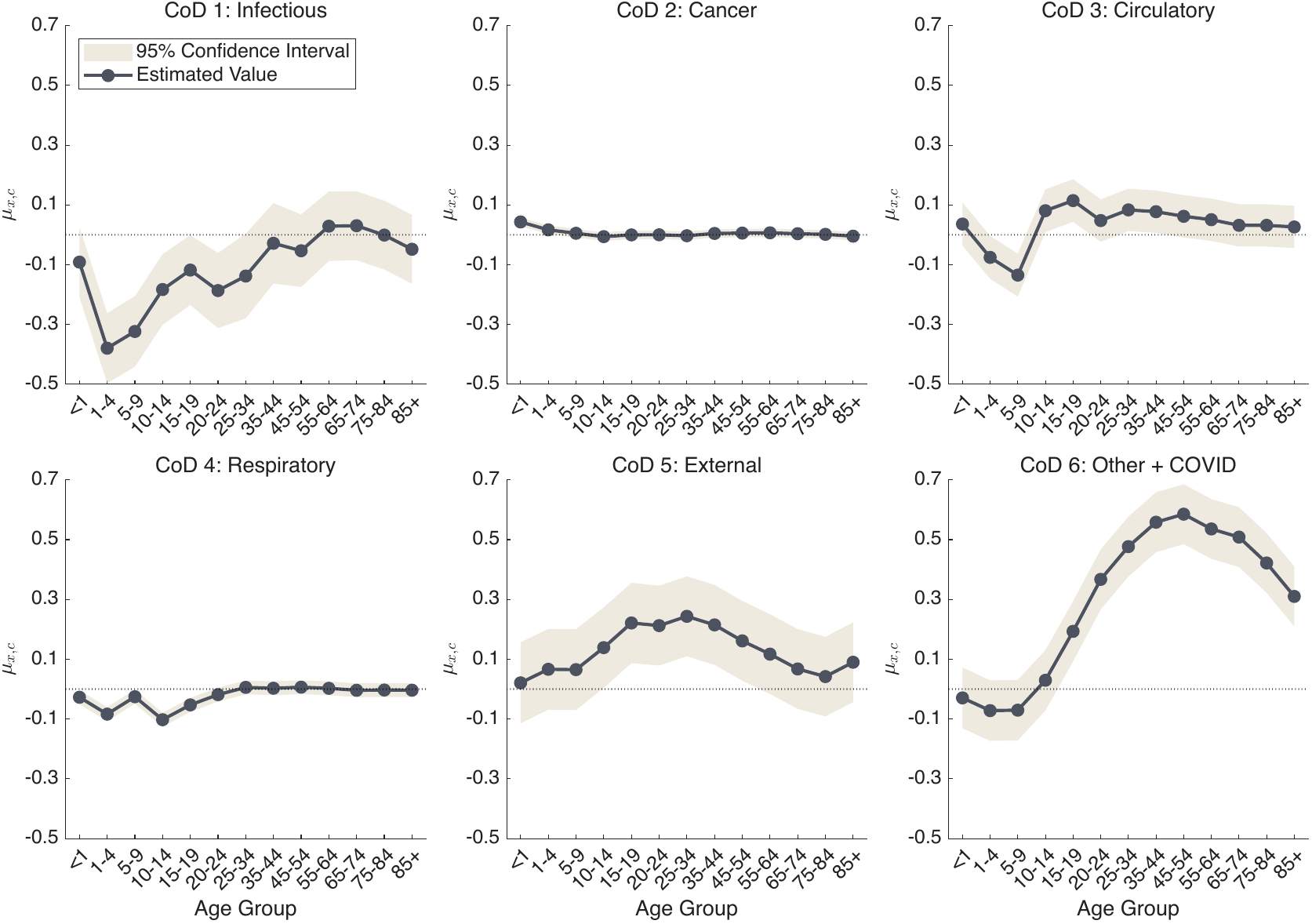}
    \caption{Estimated values of $\mu_{x,c}$ with 95\% confidence intervals from the 3WPF-CLJ model fitted to U.S. male mortality data from 1968 to 2023.}
    \label{fig:JumpMu}
\end{figure}

Figure \ref{fig:JumpMu} presents the estimated jump severity levels $\mu_{x,c}$ and their 95\% confidence intervals for the 6 causes of death. The divergent patterns across panels reflect the asymmetric impact of COVID-19 on different causes of death. Among the six CoDs, CoD 6 (Other + COVID) exhibits the highest severity level, with its age pattern of $\mu_{x,6}$ consistent with findings in \cite{Dong2020} showing that older individuals face higher COVID-19 risk. In contrast, CoD 2 (Cancer) shows only mild pandemic effects, with insignificant jump effects across all age groups except {the two youngest age groups}.

The remaining four CoDs exhibit varying effects across age groups. For CoD 1 (Infectious) and CoD 4 (Respiratory), negative values appear in young age groups, suggesting that mask regulations during COVID-19 had a protective impact on infectious and respiratory diseases for children. Notably, CoD 5 (External) shows peak values concentrated at ages 15--44, indicating that COVID-19 negatively affected young and middle-aged adults through external causes of death. This outcome aligns with findings in \cite{Czeisler2020}, where the pandemic and its associated risk mitigation activities (such as social distancing and home-working environments) posed increasing challenges to mental health.

\begin{figure}
    \centering
    \includegraphics[width=0.65\linewidth]{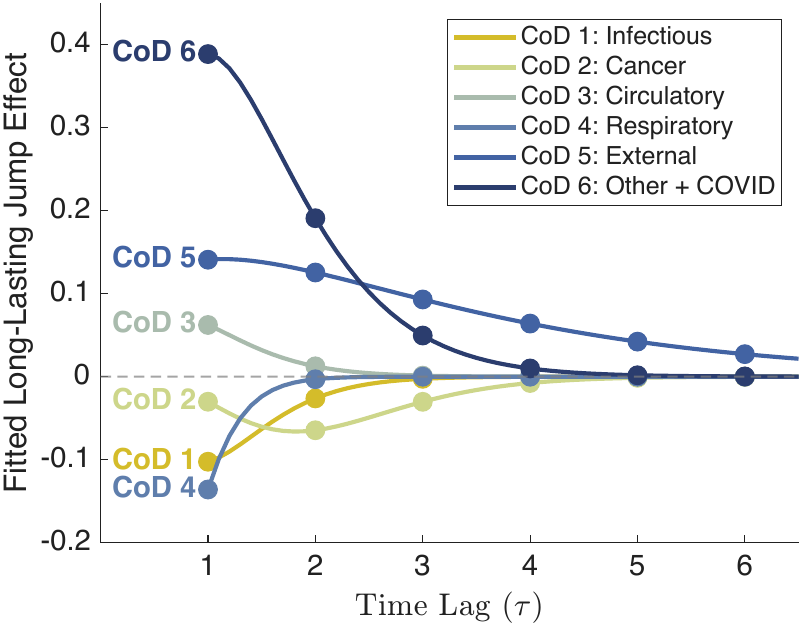}
    \caption{Fitted jump effects $\mu_{x,c} \times \pi_c(\tau)$ across causes of death from the 3WPF-CLJ model fitted to U.S. male mortality data from 1968 to 2023.}
    \label{fig:fittedCLJ}
\end{figure}

Figure \ref{fig:fittedCLJ} presents the fitted lingering jump effects $\mu_{x,c} \times \pi_c(\tau)$ for the 6 causes of death across post-pandemic years 2020--2023. The fitted patterns align closely with the empirical observations from Figure \ref{fig:LongLast}. CoD 6 (Other + COVID) exhibits the highest initial jump severity, with effects decaying rapidly over subsequent years. In contrast, CoD 5 (External) displays smaller initial severity but more persistent effects. CoD 1 (Infectious) and CoD 4 (Respiratory) show negative jump effects during the pandemic, though these benefits {largely disappear within two years}. Finally, CoD 2 (Cancer) and CoD 3 (Circulatory) exhibit only modest pandemic impacts with minimal lasting effects. {These decay patterns are estimated from the three post-jump years in the sample (2021--2023), so the values of $\pi_c(\tau)$ for $\tau>3$ are extrapolations of the fitted parametric form. Section~\ref{sec:MdlAnl_OOS} examines the first of these extrapolations against the observed data for 2024.}

{
Table \ref{tab:EstOthers} reports the estimated values and standard errors of the remaining parameters of the 3WPF-CLJ model. Although the sample contains only one major mortality shock, the observed mortality patterns across age groups and causes of death, together with their evolution from 2020 to 2023, provide useful information about the magnitude and persistence of the jump effects under the proposed model specification. The estimates of $\varphi_2$, $\varphi_3$, $\varphi_5$, and $\varphi_6$ are not statistically significant at the 5\% level. These parameters govern the cause-specific trend component, suggesting that this component is dominated by CoD 1 (Infectious) and CoD 4 (Respiratory). However, the insignificance of $\varphi_2$, $\varphi_3$, $\varphi_5$, and $\varphi_6$ does not imply weak identification of the jump effects in these causes.

For the decay parameters, $\alpha_c$ and $\beta_c$ are estimated on the log scale to ensure the positivity that is required for a gamma density function. The estimates of the magnitude parameter $\gamma_c$ are significant for all six causes at the 5\% level. However, these estimation results only characterize the observed pandemic shock, conditional on the assumed model specification. Information about the frequency of such shocks, as well as the variation of the jump magnitude across multiple catastrophic mortality events, remains limited. Accordingly, we fix $\sigma_J=0$ in the estimation and interpret $p\approx 0.0188$ (approximately one jump in 56 years of observations) with caution, since the cause-specific mortality data we use do not cover catastrophic mortality events other than the COVID-19 pandemic. 

A further limitation is the relatively short observation period available for estimating the decay parameters $\alpha_c$, $\beta_c$, and $\gamma_c$. Following the initial shock in 2020, the sample contains only three post-COVID years, 2021 to 2023. As a result, the estimates of the long-lasting jump parameters characterize the initial evolution of the pandemic effects, while their longer-term decay depends on the assumed functional form. Additional years of mortality data would allow the persistence profiles to be reassessed.
}

\begin{table}[!ht]
  \centering
  
  \caption{Parameter estimates with standard errors (S.E.) from the 3WPF-CLJ model (U.S. male mortality, 1968--2023).}
  \label{tab:EstOthers}
  \begin{tabular}{@{}lrr@{\hspace{1em}}lrr@{\hspace{1em}}lrr@{}}
    \toprule
    \multicolumn{9}{c}{\normalsize{General Parameters}} \\
    \cmidrule(lr){1-3} \cmidrule(lr){4-6} \cmidrule(lr){7-9}
    Parameter & \multicolumn{1}{c}{Estimate} & \multicolumn{1}{c}{S.E.} & Parameter & \multicolumn{1}{c}{Estimate} & \multicolumn{1}{c}{S.E.} & Parameter & \multicolumn{1}{c}{Estimate} & \multicolumn{1}{c}{S.E.} \\
    \midrule
$\varphi_1$ & {$1.0000^{\dagger}$} & {--} &
$D$ & {$-0.1701$} & {$0.0333$} &
$\log(\sigma_\eta)$ & {$-1.4107$} & {$0.1162$} \\
$\varphi_2$ & {$-0.0025$} & {$0.0217$} &
$d$ & {$0.0625$} & {$0.1465$} &
$\log(\sigma_\xi)$ & {$0.0815$} & $0.0982$ \\
$\varphi_3$ & {$0.0021$} & {$0.0217$} &
$\mathrm{logit}(p)$ & $-3.9532$ & $1.0086$ &
$\log(\sigma_J)$ & {$-\infty^{\dagger}$} & {--} \\
$\varphi_4$ & {$0.0537$} & {$0.0229$} &
 & & &
$\log(\sigma_e)$ & {$-2.7201$} & $0.0109$ \\
$\varphi_5$ & $0.0036$ & {$0.0219$} & & & & & & \\
$\varphi_6$ & {$0.0090$} & {$0.0221$} & & & & & & \\
    \midrule
    \multicolumn{9}{c}{\normalsize{Long-Lasting Effect Parameters}} \\
    \cmidrule(lr){1-3} \cmidrule(lr){4-6} \cmidrule(lr){7-9}
$\log(\alpha_c)$ & Estimate & S.E. &
$\log(\beta_c)$ & Estimate & S.E. &
$\gamma_c$ & Estimate & S.E. \\
\midrule
$c=1$ & {$1.4671$} & {$0.0268$} &
$c=1$ & {$1.3021$} & {$0.1273$} &
$c=1$ & {$0.1246$} & {$0.0192$} \\
$c=2$ & {$1.8366$} & {$0.0146$} &
$c=2$ & {$1.0635$} & {$0.0525$} &
$c=2$ & {$-0.1229$} & {$0.0197$} \\
$c=3$ & {$1.2006$} & {$0.0353$} &
$c=3$ & {$1.1674$} & {$0.2311$} &
$c=3$ & {$0.9662$} & {$0.1327$} \\
$c=4$ & {$1.2565$} & {$0.0131$} &
$c=4$ & {$1.7082$} & {$0.0398$} &
$c=4$ & {$3.5941$} & {$0.2862$} \\
$c=5$ & {$0.4932$} & {$0.0974$} &
$c=5$ & {$-0.5837$} & {$0.1066$} &
$c=5$ & {$5.0137$} & {$0.6474$} \\
$c=6$ & {$1.1684$} & {$0.0135$} &
$c=6$ & {$0.8098$} & {$0.0284$} &
$c=6$ & {$0.9263$} & {$0.0353$} \\
\bottomrule
\end{tabular}
\par\medskip
\begin{minipage}{\linewidth}
\footnotesize
{\textit{Notes:} 
$^{\dagger}$Fixed by constraint at $\varphi_1=1$ and $\sigma_J=0$
(i.e., $\log(\sigma_J)=-\infty$).}
\end{minipage}
\end{table}

\subsection{Robustness}  \label{sec:MdlAnl_Robust}
This subsection examines the robustness of the parameter estimates under two scenarios: (1) varying the calibration window, and (2) removing the data from COVID years.

\subsubsection{Robustness to Calibration Windows} \label{sec:MdlAnl_Robust_Window}

\begin{figure}[!ht]
    \centering
    \includegraphics[width=\textwidth]{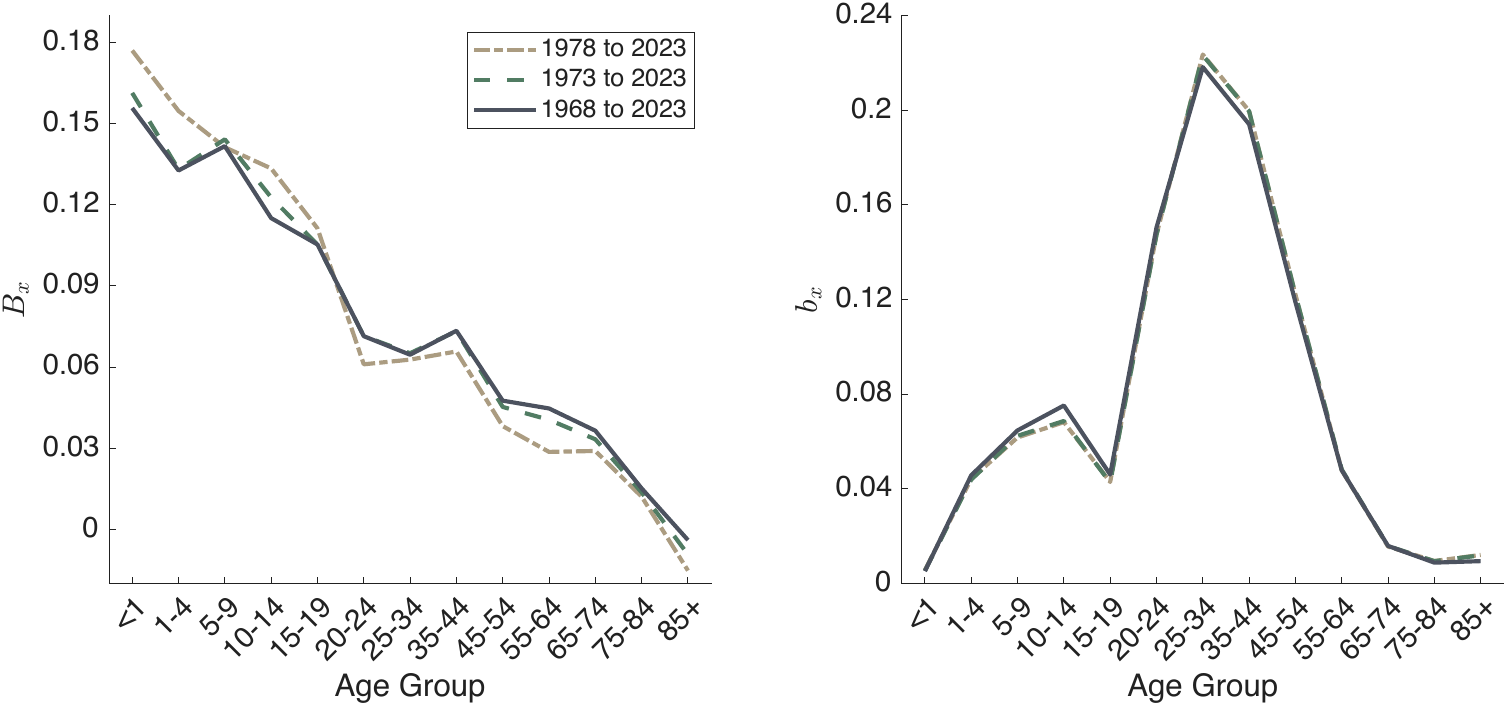}
    \caption{Estimated values of $B_x$ and $b_x$ from the 3WPF-CLJ model fitted to U.S. male mortality data across three calibration windows {with fixed ending year at 2023}: 1968--2023, 1973--2023, and 1978--2023.}
    \label{fig:1968vs1973vs1978}
\end{figure}

We first study the robustness of the proposed 3WPF-CLJ model across different sample periods by fixing the ending year at 2023 and varying the starting year among 1968, 1973, and 1978. The model is fitted to these three calibration windows without any adjustment. Because the COVID-19 pandemic is included in all three calibration windows, the model's estimated jump effect remains nearly identical. Figure \ref{fig:1968vs1973vs1978} shows the estimated values of $B_x$ and $b_x$ across the three windows. The pattern of $b_x$ remains stable while the downward slope of $B_x$ is largely preserved across all calibration windows, with only minor variations across age groups.

\subsubsection{Robustness to Pandemic Data} \label{sec:MdlAnl_Robust_Pandemic}

We next examine the robustness of non-jump-related parameters; specifically, $B_x$ and $D$ for the general mortality trend, and $\varphi_c$, $b_x$ and $d$ for the cause-specific mortality dynamics. We compare two calibration windows with the same starting year but different ending years: 1968--2019 and 1968--2023. For the first calibration window (1968--2019), COVID-19 had not yet occurred. Therefore, when calibrating the model, we disable the jump effects by imposing $N_{1968}=N_{1969}=\cdots=N_{2019}=0$ in equation \eqref{eqn:zxti simp}, which yields
\begin{equation} \label{eqn:zxti no Nt}
Z_{x,t,c} = B_x(D+\eta_t)+\varphi_cb_x(d+\xi_t)+e_{x,t,c}-e_{x,t-1,c}.
\end{equation}
For the second calibration window (1968--2023), the proposed model described in equation \eqref{eqn:zxti simp} is used without modification.

\begin{figure}[!ht]
    \centering
    \includegraphics[width=\textwidth]{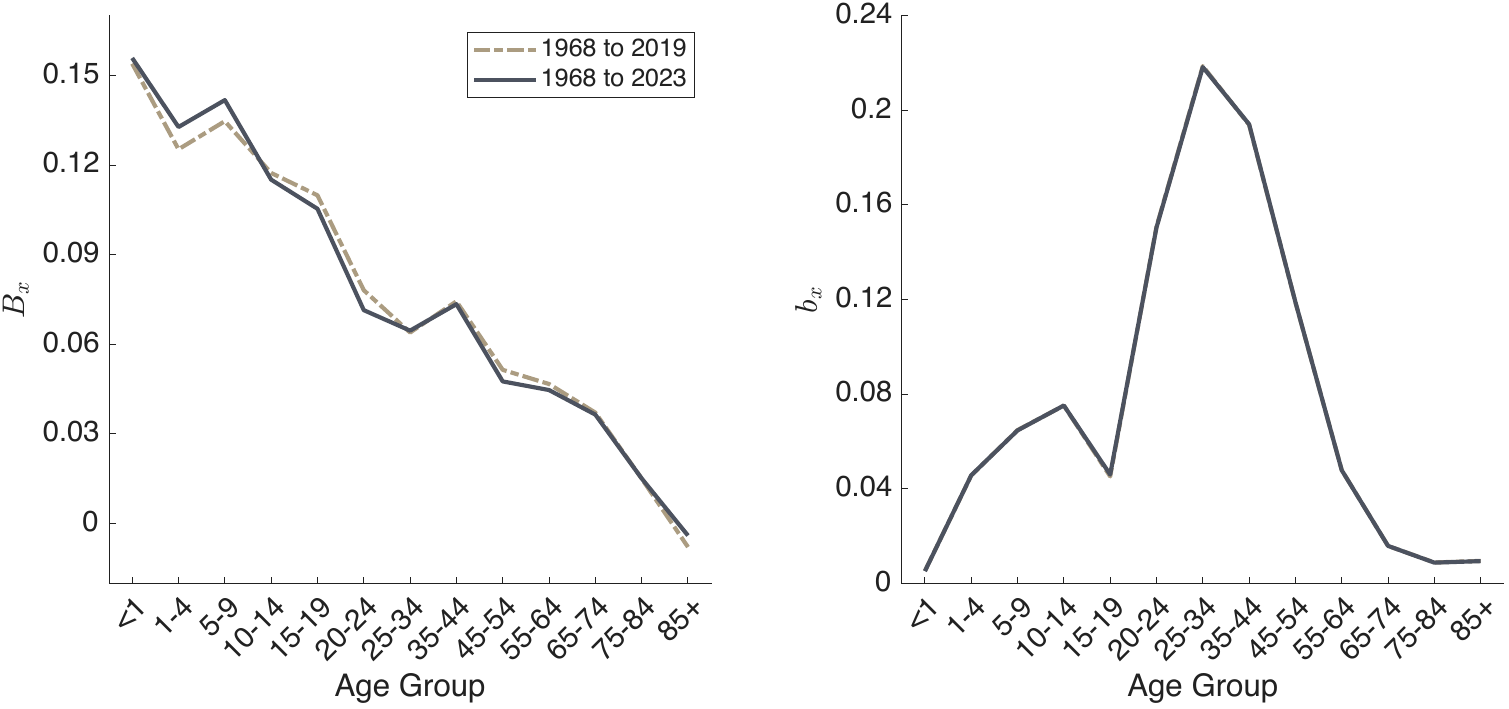}
    \caption{Estimated values of $B_x$ and $b_x$ from the 3WPF-CLJ model fitted to U.S. male mortality data across two calibration windows  {with fixed starting year at 1968}: 1968--2019 and 1968--2023.}
    \label{fig:2019vs2023}
\end{figure}

Figure \ref{fig:2019vs2023} compares the estimated values of $B_x$ and $b_x$ across the two calibration windows. The results reveal that all non-jump-related parameters are remarkably similar across both calibration windows. This outcome implies that the jump component in the 3WPF-CLJ model successfully absorbs pandemic effects, thereby isolating the non-jump-related parameters from the COVID-19 pandemic.

\subsection{Jump Effects} \label{sec:MdlAnl_Jump}
This subsection demonstrates the necessity of (1) incorporating jump effects into the model and (2) allowing these effects to vary by cause of death.

\subsubsection{Importance of the Jump Component} \label{sec:MdlAnl_Jump_Comp}

We first examine the importance of having a jump component in the model by comparing the 3WPF-CLJ model with its no-jump counterpart fitted to the complete dataset (1968--2023). The no-jump model corresponds to equation \eqref{eqn:zxti no Nt} with $N_t=0$ for all $t$. As discussed in Section \ref{sec:datamotiv_heterogeneity}, the COVID-19 pandemic affects different causes of death asymmetrically. When the jump component is omitted, these asymmetric effects will be absorbed by other model parameters.

\begin{figure}[!ht]
    \centering
    \includegraphics[width=\textwidth]{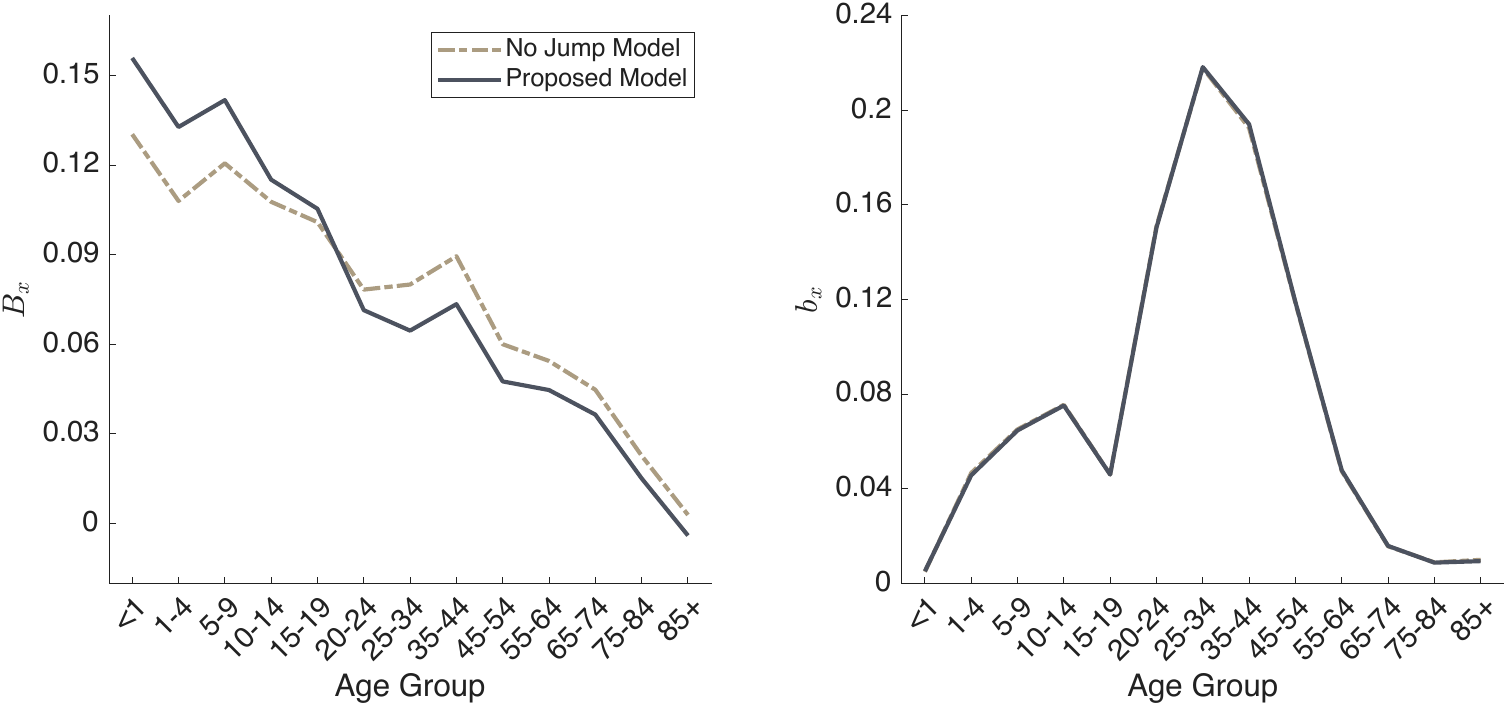}
    \caption{Estimated values of $B_x$ and $b_x$ from the 3WPF-CLJ model and the 3WPF (no-jump) model fitted to U.S. male mortality data from 1968 to 2023.}
    \label{fig:JumpvsNoJump}
\end{figure}

Figure \ref{fig:JumpvsNoJump} presents the estimated values of $B_x$ and $b_x$ from the two models. The flatter shape of $B_x$ under the no-jump model compared to the 3WPF-CLJ model reflects the aggregation of the asymmetric effects across causes. Without the jump component, this distorted pattern of $B_x$ would be incorrectly treated as a long-term structural change, leading modelers to generate biased forecasts of general mortality trends. In contrast, the estimated $b_x$ remains largely unchanged between the two models.

\subsubsection{Importance of Cause-Specific Jumps} \label{sec:MdlAnl_Jump_Cause}

\begin{figure}[!ht]
    \centering
    \includegraphics[width=0.7\textwidth]{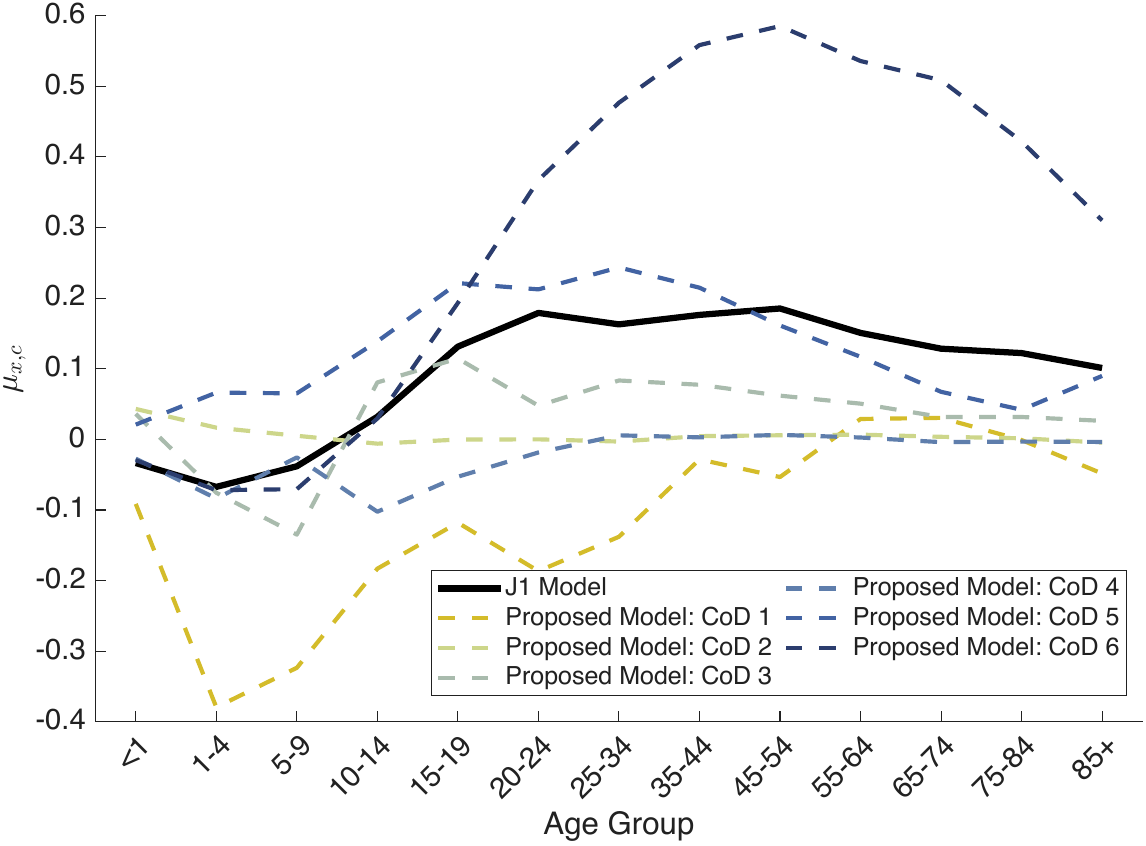}
    \caption{Estimated jump effects from the 3WPF-CLJ model (age- and cause-specific) and the J1 model (age-specific only) fitted to U.S. male mortality data from 1968 to 2023.}
    \label{fig:J1vsProposed}
\end{figure}

Lastly, we investigate the importance of allowing cause-specific jumps by comparing the proposed 3WPF-CLJ model with the J1 model from \cite{LiuLi2015}\footnote{A brief review of the J1 model is provided in Appendix \ref{app:review}.}, which permits age-specific but not cause-specific mortality jumps. Figure \ref{fig:J1vsProposed} compares the estimated mean of the jump effects: $\mu_{x,c}$ for the 3WPF-CLJ model and $\mu_{x}$ for the J1 model.

In the J1 model, the absence of cause-specific heterogeneity forces the estimated jump pattern to reflect the pandemic's overall impact on mortality. Consequently, the jump severity from the J1 model resembles a weighted average of the cause-specific jump effects from the proposed model. This averaging leads to systematic biases: overestimation of jump effects for young age groups in CoD 1 (Infectious) and CoD 4 (Respiratory), and underestimation of jump effects for old age groups in CoD 6 (Other + COVID). These findings underscore the importance of incorporating cause-specific heterogeneity when modeling pandemic mortality shocks.

{

\subsection{Parsimony in the Long-Lasting Jump Model} \label{sec:MdlAnl_Parsimony}

The proposed 3WPF-CLJ model assumes cause-specific parameters to capture the long-lasting effects of COVID-19. In this subsection, we study the parsimony of the proposed model by comparing various 3WPF-CLJ model variants, where some (or all) of the long-lasting jump parameters (or the decay parameters) are shared across causes.

The results of the comparison are provided in Table \ref{tab:llj_parameter_sensitivity}. Among all the candidate models, the baseline model with all cause-specific long-lasting jump parameters scores the highest log-likelihood and the lowest AIC. With cause-specific long-lasting jump parameters, the post-COVID dynamics attributable to different causes are captured more adequately than in the variants with common long-lasting parameters. The benefit of this richer structure is supported by the AIC, which penalizes the log-likelihood by the number of effective parameters. Since cause-specific long-lasting effects are the main focus of this paper, we prefer the 3WPF-CLJ model with cause-specific long-lasting jump effects.

It is also worth noting in Table \ref{tab:llj_parameter_sensitivity} that the model variant with a common shape parameter $\alpha$, cause-specific rate parameters $\beta_c$, and a common magnitude parameter $\gamma$ scores the lowest BIC. Both AIC and BIC balance the fit against the complexity of the model, but BIC places more weight on complexity. Therefore, if a more parsimonious model is the main objective, the variant with a common $\alpha$, cause-specific $\beta_c$, and a common $\gamma$ is also a good candidate. 
 
\begin{table}[htbp]
\centering

\caption{Sensitivity analysis of common long-lasting jump parameters across causes of death in 3WPF-CLJ Model using the U.S. male data from 1968 to 2023.}
\label{tab:llj_parameter_sensitivity}
\begin{tabular}{c|c|r|r|r}
\toprule
& Number of & & & \\
Common parameters & Effective Parameters & Log-likelihood & AIC & BIC \\
\midrule
None & 131 & 4548.38 & $\mathbf{-8834.76}$ & $-8001.07$ \\
$\alpha$                & 126 & 4542.45 & $-8832.91$ & $-8031.04$ \\
$\beta$                 & 126 & 4543.18 & $-8834.36$ & $-8032.49$ \\
$\gamma$                & 126 & 4535.08 & $-8818.17$ & $-8016.30$ \\
$\alpha$ and $\beta$          & 121 & 4520.25 & $-8798.50$ & $-8028.45$ \\
$\alpha$ and $\gamma$         & 121 & 4526.72 & $-8811.43$ & $\mathbf{-8041.38}$ \\
$\beta$ and $\gamma$          & 121 & 4514.60 & $-8787.19$ & $-8017.14$ \\
$\alpha,\beta$ and $\gamma$   & 116 & 4503.83 & $-8775.66$ & $-8037.43$ \\
\bottomrule
\end{tabular}

\medskip
\begin{minipage}{\linewidth}
\footnotesize
\textit{Notes:}
Parameters not listed as common are cause-specific. Parameter counts deduct the three constraints $\sum_x B_x=1$, $\sum_x b_x=1$, and $\varphi_1=1$. $\mathrm{AIC}=-2\ell+2k$ and $\mathrm{BIC}=-2\ell+k\log(n)$, where $k$ is the number of effective parameters, $\ell$ is the log-likelihood, and $n=4290$ is the number of observations.
\end{minipage}
\end{table}

An alternative way to allow partial parameter sharing is to apply parameter regularization. To be more specific, we can consider the following objective function with additional penalty terms:
\begin{equation} \label{eqn: loglik-regularization}    
l^{\ast}(\vec{\theta}) = l(\vec{\theta})-\lambda_\alpha P_\alpha  - \lambda_\beta P_\beta - \lambda_\gamma P_\gamma.
\end{equation}
where $l(\vec{\theta})$ is the log-likelihood defined in equation \eqref{eqn: loglik}, and 
\[
P_\alpha = \sum_{1\leq i < j \leq 6}\frac{|\alpha_i-\alpha_j|}{s_\alpha}, \quad P_\beta = \sum_{1\leq i < j \leq 6}\frac{|\beta_i-\beta_j|}{s_\beta}, \quad P_\gamma =  \sum_{1\leq i < j \leq 6}\frac{|\gamma_i-\gamma_j|}{s_\gamma} 
\]
represent L1 penalties\footnote{L1 penalty has the ability to enforce equality into the long-lasting parameters and is therefore preferred over L2 penalty.} on differences across causes, adjusted by their own scale parameters to ensure the magnitudes of each parameter do not inadvertently dominate the overall shared penalty. The scale parameter can be calculated from the root mean squares. When regularization is used, the penalty parameters can be selected by considering a grid search over a range of possible values. 
However, the estimation of the model can only be achieved numerically since the partial derivatives of the penalized log-likelihood have no explicit form. We would leave this topic for future study. 

\subsection{Sensitivity to the Inclusion of COVID-19 Deaths in CoD 6} \label{sec:MdlAnl_Sensitivity}

In this paper, CoD 6 is defined as a heterogeneous residual category that includes: 1) deaths recorded with COVID-19 as the underlying cause, and 2) all other deaths that are not included in the first five categories. Consequently, its estimated jump component may reflect both recorded COVID-19 mortality and changes among the remaining causes. To examine the sensitivity of the results to this aggregation, we remove recorded COVID-19 deaths from CoD 6 and refit the 3WPF-CLJ model using the same U.S. male mortality sample from 1968 to 2023.

Figure \ref{fig:cod6-covid-exclusion} compares the fitted percentage effects of a jump occurring in 2020 under the two specifications, that is, with or without COVID-19 deaths. For each fitted model, the percentage effect at age group $x$ and lag $\tau$ is calculated as
\[
100 \times \left(\exp(\mu_{x,6}\pi_6(\tau)) -1\right),\quad \quad \text{ where } \tau = 0, 1, 2, 3. 
\]
This quantity measures the mortality-level contribution of the fitted jump relative to the corresponding trajectory when the jump component is suppressed. We can observe from the figure that excluding COVID-19 deaths substantially reduces the estimated positive jump effects among adult age groups. Nevertheless, the remaining effects are not uniformly close to zero. They are concentrated around ages 25–44, peak in year 2021, and decline thereafter. For example, among individuals aged 35–44, the COVID-excluded estimate is approximately 35\% in 2021 and remains around 15\% in 2023. The COVID-included estimates are initially much larger but decline more rapidly, so that the excluded estimates exceed them at some working ages by year 2023. 

The comparison reveals differences in both the magnitude and the persistence of the fitted effects. It shows that the inclusion of recorded COVID-19 deaths strongly influences the estimated jump profile of the residual category. The result also indicates that removing those deaths does not make the remaining CoD 6 jump component negligible in magnitude. Therefore, the appropriate composition of CoD 6 depends on the research objective. If the main objective is to analyze mortality changes among the residual non-COVID causes following the onset of the pandemic, deaths recorded with COVID-19 as the underlying cause should be excluded from CoD 6 before model estimation. This exclusion reveals the fitted jump effects associated specifically with the remaining causes, which may otherwise be obscured by the large contribution of recorded excess COVID-19 mortality. 

\begin{figure}[th!]
    \centering
    \includegraphics[width=0.9\textwidth]{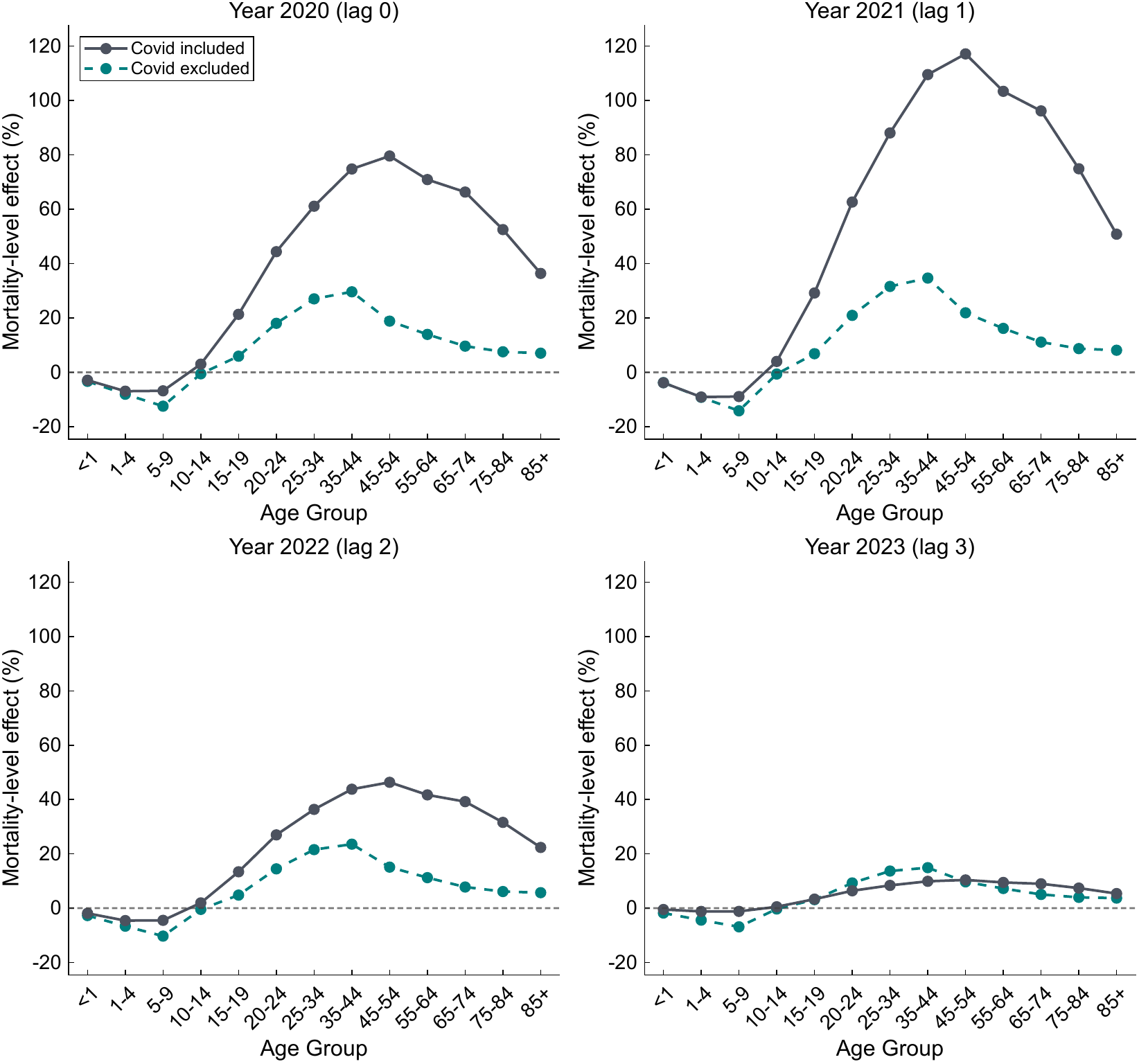}
    \caption{Estimated CoD 6 mortality effects of a 2020 jump, with and without COVID-19 deaths, using the U.S. male population from 1968-2023.}
    \label{fig:cod6-covid-exclusion}
\end{figure}
}

{
\subsection{Out-of-Sample Analysis} \label{sec:MdlAnl_OOS}

We now conduct an out-of-sample analysis. Using the model fitted to data up to 2023, we forecast beyond 2023 and examine the forecasting performance against the observed mortality in 2024, as well as the plausibility of the long-term projection.

We first examine the forecasting performance on all age- and cause-specific death rates in 2024. We consider two one-year-ahead forecasts, both starting from the observed rates of 2023. The first is a benchmark without further decay, in which the jump effect embedded in the 2023 rates is carried forward unchanged. The second is the full 3WPF-CLJ model, in which the lingering jump effect continues to fade between 2023 and 2024. Table~\ref{tab:OOS2024} reports the root mean squared error (RMSE) of the log death rates and the mean absolute percentage error (MAPE) of the death rates. To aggregate the errors across ages and causes, we consider three weightings: (1) equal weights, (2) exposure weights based on the 2023 exposures, and (3) death weights based on the 2023 death counts. The further decay reduces the errors under all three weightings, most clearly under death weights, where the RMSE falls from 0.060 to 0.048. 

We next examine the long-term projection in terms of life expectancy at age 35. The projection is based on 50,000 paths simulated from the fitted model, with random period effects, random future jumps and the error term. For each simulated path, we sum the projected cause-specific death rates across the six causes and compute life expectancy from the resulting all-cause rates. Figure~\ref{fig:OOS_e35} shows the observed life expectancy at age 35, its projection from 2024 to 2040 and the 90\% predictive interval.\footnote{The lower edge of the interval drops in 2026, because from then on more than 5\% of the simulated paths contain at least one new jump, which lowers life expectancy substantially.} In 2024, the observed value of 43.58 years lies inside the interval $(42.86, 43.67)$. Over the longer term, the projection is also plausible. The projected gain of 0.10 years per year over 2023--2030 lies between the observed averages of 0.12 years over 2000--2019 and 0.03 years over 2010--2019, and it slows to about 0.06 years per year after 2030.

\begin{table}[ht!]
\centering
\caption{Out-of-sample forecasting performance of the fitted 3WPF-CLJ model for 2024.}
\label{tab:OOS2024}
\begin{tabular}{l rrr rrr}
\toprule
 & \multicolumn{3}{c}{RMSE} & \multicolumn{3}{c}{MAPE (\%)} \\
\cmidrule(lr){2-4} \cmidrule(lr){5-7}
Forecast & Equal & Exposure & Death & Equal & Exposure & Death \\
\midrule
Without further decay & 0.0914 & 0.0908 & 0.0604 & 6.95 & 6.83 & 4.22 \\
With further decay    & \bf 0.0898 & \bf 0.0825 & \bf 0.0476 & \bf 6.65 & \bf 6.06 & \bf 3.53 \\
\bottomrule
\end{tabular}
\end{table}

\begin{figure}[th!]
    \centering
    \includegraphics[width=0.65\textwidth]{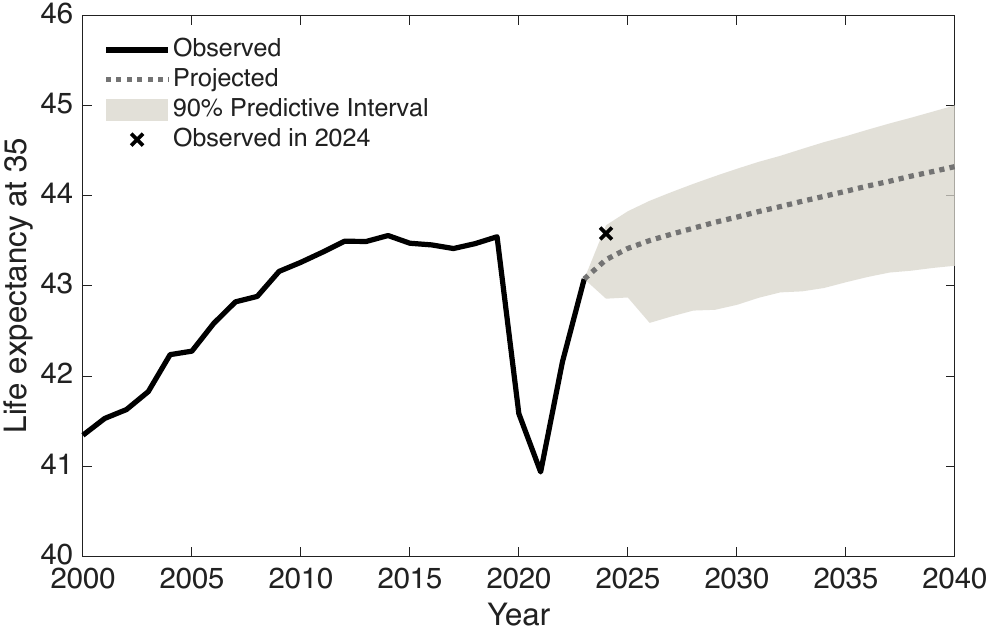}
    \caption{Projected life expectancy at age 35 from the fitted 3WPF-CLJ model.}
    \label{fig:OOS_e35}
\end{figure}
}

\section{Risk Management Applications} \label{sec:Application}

This section demonstrates the application of the 3WPF-CLJ model in life insurance valuation and risk management. We first present the valuation framework for stylized life insurance products and their distributional characteristics under the 3WPF-CLJ model. We then compare our model to existing models under a natural hedging strategy. Finally, we conduct scenario analysis to assess how alternative dynamics of mortality jumps affect risk profiles and hedging effectiveness.

\subsection{Stylized Products} \label{sec:ApplicationProducts}
We consider two stylized products to demonstrate the impact of age- and cause-specific lingering jump effects on liability distributions. Let $t_0$ be the valuation date, assumed to be the end of the sample period (i.e., the end of year 2023). The survival probability for an individual aged $x$ at time $t_0$ to reach time $t_0+T$ is given by
\[
S_{x,t_0}(T) = \exp \left( - \sum_{s=1}^{T} \sum_{c=1}^{6} m_{x+s-1,t_0+s,c} \right).
\]
Our proposed model generates stochastic projections of mortality rates $m_{x,t_0,c}$ to obtain the empirical distribution of $S_{x,t_0}(T)$.

The first product is a $T$-year term life annuity issued to an individual aged $x$ at time $t_0$, with a payment of \$1 at the end of each year conditional on survival. The actuarial present value of this annuity is
\[
\mathcal{A} = \sum_{t=1}^{T} (1+r)^{-t} S_{x,t_0}(t),
\]
where $r$ is the constant interest rate for discounting future cash flows. The second product is a $T$-year term life insurance issued to an individual aged $x$ at time $t_0$, with a death benefit of \$1 payable at the end of the year of death. The actuarial present value of this insurance is
\[
\mathcal{I} = \sum_{t=0}^{T-1} (1+r)^{-(t+1)} S_{x,t_0}(t) (1 - S_{x+t,t_0+t}(1)).
\]

Under the 3WPF-CLJ model, jumps affect mortality heterogeneously across ages and causes, leading to asymmetric distributions of $\mathcal{I}$ and $\mathcal{A}$. For life insurance, a future pandemic will increase death probabilities, shifting the distribution of $\mathcal{I}$ to the right, while the opposite is true for life annuities, shifting the distribution of $\mathcal{A}$ to the left. Table~\ref{tab:product_dist} presents distributional characteristics of mean-adjusted present values (i.e., $\mathcal{A} - \mbox{E}[\mathcal{A}]$ and $\mathcal{I} - \mbox{E}[\mathcal{I}]$) for a 30-year deferred annuity issued at age 35 with 30-year payment term and a 30-year term life insurance issued at age 35, with face values determined by setting $\mbox{E}[\mathcal{A}] = \mbox{E}[\mathcal{I}] = 100$.

\begin{table}[ht!]
\centering
\caption{Distributional characteristics of mean-adjusted actuarial present values for the annuity, life insurance, and hedged portfolio under the estimated 3WPF-CLJ model.}
\label{tab:product_dist}
\begin{tabular}{lcccccc}
\toprule
Product & CTE$_{5\%}$ & VaR$_{5\%}$ & VaR$_{95\%}$ & CTE$_{95\%}$ & S.D. & Skewness \\
\midrule
Annuity & {$-3.49$} & {$-2.73$} & {$2.61$} & {$3.22$} & {$1.62$} & {$-0.15$} \\
Insurance & {$-8.23$} & {$-6.59$} & {$7.41$} & {$9.51$} & {$4.29$} & {$0.28$} \\
Portfolio & {$-1.19$} & {$-0.93$} & {$0.92$} & {$1.20$} & {$0.57$} & {$0.02$} \\
\bottomrule
\end{tabular}
\end{table}

Table~\ref{tab:product_dist} quantifies the asymmetric impact of age- and cause-specific mortality jumps. The annuity exhibits negative skewness of {$-0.15$}, with left-tail risk exceeding right-tail risk in absolute value, reflecting that mortality spikes reduce liability values. Conversely, the insurance product displays positive skewness of {$0.28$}, with right-tail measures dominating left-tail measures in absolute value due to increased death benefit liabilities from pandemics. The insurance product also exhibits substantially higher standard deviation ({$4.29$ versus $1.62$}), suggesting greater sensitivity to mortality shocks.

\subsection{Natural Hedging} \label{sec:ApplicationHedging}

The opposing tail risks and volatility differences shown in Table~\ref{tab:product_dist} motivate a natural hedging strategy that combines insurance and annuity liabilities to offset their asymmetric mortality exposures. Natural hedging exploits the negative correlation between mortality and longevity risks to construct diversified portfolios \citep{cox2007natural}. Various approaches have been proposed to calibrate the optimal hedging ratio \citep[see, e.g.,][]{wang2010optimal, lin2014applications, luciano2017single, cupido2024spatial}. We adopt the variance minimization approach of \cite{cairns2014longevity} to determine the optimal portfolio mix.

The present value of a combined portfolio is
\begin{equation*} \label{eq:vAuI}
\mathcal{P} = \omega\mathcal{A} + (1-\omega)\mathcal{I},
\end{equation*}
where $\omega \in [0,1]$ represents the proportion of annuity liabilities in the portfolio. The optimal weight $\omega^*$ is determined by minimizing the variance of $\mathcal{P}$, which yields
\begin{equation*} \label{eq:omega}
\omega^* = \frac{\text{Var}(\mathcal{I}) - \text{Cov}(\mathcal{A},\mathcal{I})}{\text{Var}(\mathcal{A}) + \text{Var}(\mathcal{I}) - 2\text{Cov}(\mathcal{A},\mathcal{I})}.
\end{equation*}
This approach balances the products' relative volatility and correlation structure to achieve maximum risk reduction.

\begin{figure}[ht!]
    \centering
    \includegraphics[width=0.8\linewidth]{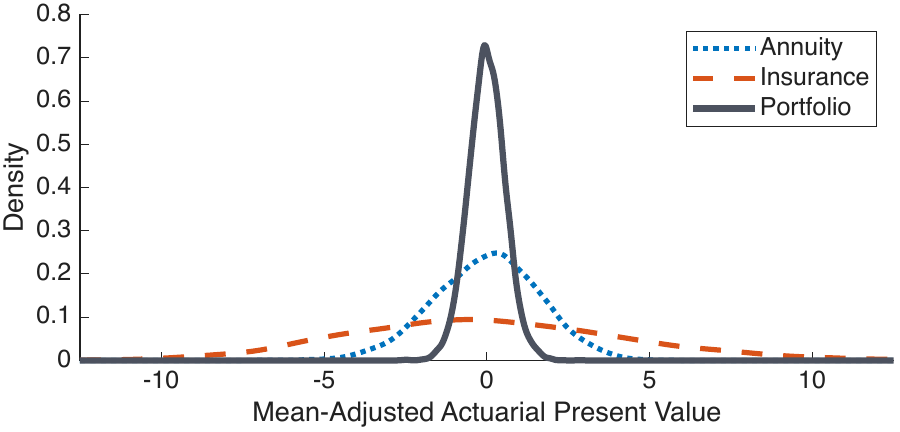}
    \caption{Probability density functions of mean-adjusted actuarial present values for the annuity, life insurance, and hedged portfolio under the estimated 3WPF-CLJ model.}
    \label{fig:Baseline}
\end{figure}

Based on the two products specified in Section~\ref{sec:ApplicationProducts}, the optimal portfolio allocation is $\omega^* = 0.74$, placing 74\% weight on annuity liabilities and 26\% on insurance liabilities. As shown in Table~\ref{tab:product_dist} and Figure~\ref{fig:Baseline}, the hedged portfolio achieves substantial risk reduction. Standard deviation declines to {$0.57$} compared to {$1.62$} for the annuity and {$4.29$} for the insurance product, while skewness of {$0.02$} indicates that opposing tail risks effectively neutralize each other. The narrower symmetric distribution of the portfolio in Figure~\ref{fig:Baseline} visually confirms the effectiveness of natural hedging even when age- and cause-specific mortality jumps are present.

We now compare the proposed model with alternative specifications to examine how jump dynamics affect optimal hedge ratios. The analysis focuses on models with varying levels of jump effect specificity: \begin{itemize}
    \item \textbf{Proposed 3WPF-CLJ Model:} Age- and cause-specific jump effects.
    \item \textbf{J1 Model \citep{LiuLi2015}:} Age-specific jump effects only.
    \item \textbf{CC Model \citep{ChenCox2009}:} Aggregate jump on the common period effect.
\end{itemize}
The CC and J1 models follow equation \eqref{eqn:ChenCox2009} and \eqref{eqn: Liu and Li 2015} respectively, and are reviewed in more detail in Appendix~\ref{app:review}. Table~\ref{tab:ScenarioWeight} reports optimal weights and portfolio characteristics under each model specification. To isolate the impact of misspecified hedge ratios, all portfolio statistics are computed using mortality projections from the proposed 3WPF-CLJ model, while hedge ratios $\omega^*$ are calibrated under each alternative model's specification.

\begin{table}[ht!]
\centering
\caption{Optimal weights and distributional characteristics under the proposed 3WPF-CLJ model, the J1 model, and the CC model.}
\label{tab:ScenarioWeight}
\begin{tabular}{lccccccc}
\toprule
Model & $\omega^*$ & CTE$_{5\%}$ & VaR$_{5\%}$ & VaR$_{95\%}$ & CTE$_{95\%}$ & S.D. & Skewness \\
\midrule
Proposed & $0.74$ & {$-1.19$} & {$-0.93$} & {$0.92$} & {$1.20$} & {$0.57$} & {$0.02$} \\
J1 & $0.65$ & {$-1.43$} & {$-1.17$} & {$1.34$} & {$1.77$} & {$0.76$} & {$0.41$} \\
CC & $0.67$ & {$-1.29$} & {$-1.04$} & {$1.17$} & {$1.55$} & {$0.67$} & {$0.36$} \\
\bottomrule
\end{tabular}
\end{table}

The J1 and CC models produce moderately lower hedge ratios at $0.65$ and $0.67$, respectively, compared to $0.74$ for the proposed model. These seemingly modest differences translate into meaningful increases in portfolio risk. Portfolio standard deviation increases from {$0.57$} to {$0.76$} under the J1 model and to {$0.67$} under the CC model. Both alternatives also exhibit substantially higher positive skewness ({$0.41$} for the J1 model and {$0.36$} for the CC model versus {$0.02$} for the proposed model). These results demonstrate that age- and cause-specific jump modeling improves hedging effectiveness by more accurately capturing heterogeneous pandemic impacts across age groups and causes of death.

\subsection{What-If Analysis}\label{sec:ApplicationWhatif}

This subsection demonstrates the flexibility of the 3WPF-CLJ model in evaluating alternative mortality dynamics through what-if analysis. By modifying the jump-related parameters, we simulate four distinct future mortality scenarios and assess their implications under the natural hedging framework established in Section~\ref{sec:ApplicationHedging}. The four scenarios are constructed as follows:
\begin{itemize}
    \item \textbf{Scenario I (No future pandemics):} No new jumps occur in the future 
    ($p = 0$).
    \item \textbf{Scenario II (Endemic regime):} Jump frequency is quadrupled 
    ($p = 4\hat{p}$) and jump severity is halved 
    ($\mu_{x,c}^{(J)} = 0.50\,\hat{\mu}_{x,c}^{(J)}$), from the baseline in Section \ref{sec:MdlAnl_Est}.
    \item \textbf{Scenario III (Medical breakthrough):} A single permanent 50\% reduction in cancer mortality (CoD~2) across all ages with annual probability $p = 1\%$.
    \item \textbf{Scenario IV (Catastrophic event):} Recurring transitory shocks that increase external cause mortality (CoD~5) tenfold among working-ages (35--64) with annual probability $p = 1\%$.
\end{itemize}
These scenarios represent plausible mortality futures, from the complete absence of future shocks to persistent endemic events, sudden longevity improvements, and concentrated catastrophic mortality shocks. In all scenarios, the long-lasting effects of COVID-19 are retained in the forecasts, and the baseline natural hedge ratio $\omega^* = 0.74$ is held fixed.

\begin{figure}[ht!]
    \centering
    \includegraphics[width=\linewidth]{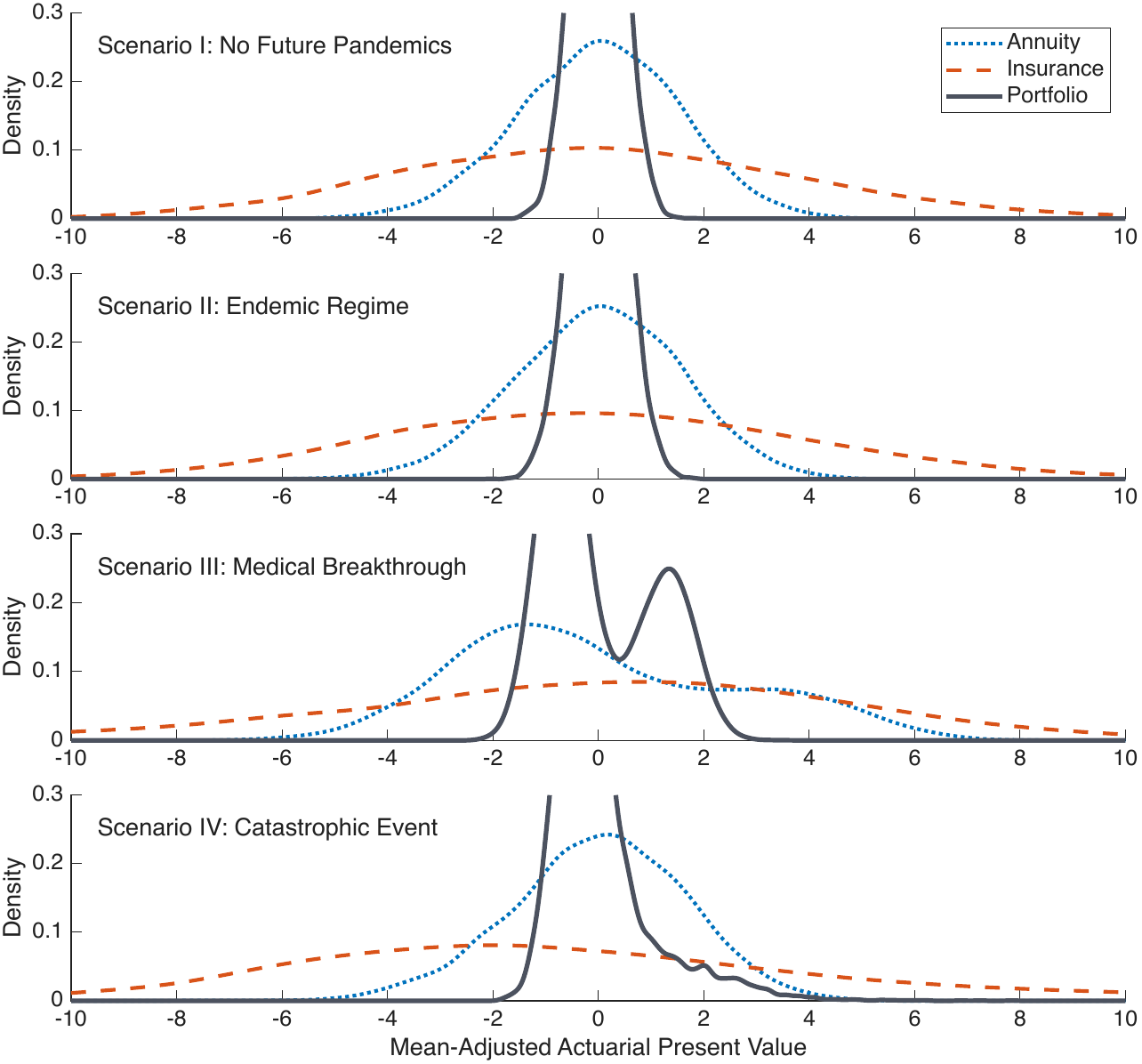}
    \caption{Probability density functions of mean-adjusted actuarial present values for the annuity, life insurance, and hedged portfolio under what-if mortality scenarios.}
    \label{fig:WhatIf}
\end{figure}

Table~\ref{tab:WhatIfStats} reports the standard deviation and skewness of the annuity, insurance, and hedged portfolio under each scenario, and Figure~\ref{fig:WhatIf} illustrates the corresponding probability density functions of their mean-adjusted actuarial present values. In Scenario~I, where future jumps are entirely eliminated, standard deviation falls across all three products and skewness is greatly reduced. Most notably, the hedged portfolio's standard deviation declines from {$0.57$} to {$0.43$}, and the insurance product's skewness drops from {$0.28$} to {$0.16$}. Scenario~II, which assumes an endemic regime of higher-frequency but lower-severity shocks, produces an intermediate risk profile. Standard deviation rises only modestly relative to Scenario~I, and skewness stays similarly subdued. This suggests that a regime of frequent mild shocks generates materially less tail risk than the single severe COVID-19 event embedded in the baseline.

\begin{table}[ht!]
  \centering
  \caption{Standard deviation and skewness of mean-adjusted actuarial present values for the annuity, life insurance, and hedged portfolio under what-if mortality scenarios.}
  \label{tab:WhatIfStats}
  \begin{tabular}{@{}l@{\hspace{1.5em}}rrr@{\hspace{1.5em}}rrr@{}}
    \toprule
    & \multicolumn{3}{c}{Standard Deviation} & \multicolumn{3}{c}{Skewness} \\
    \cmidrule(lr){2-4} \cmidrule(lr){5-7}
    Scenario & Annuity & Insurance & Portfolio & Annuity & Insurance & Portfolio \\
    \midrule
    Baseline     & {$1.62$} & {$4.29$} & {$0.57$} & {$-0.15$} & {$ 0.28$} & {$ 0.02$} \\
    Scenario I   & {$1.54$} & {$3.85$} & {$0.43$} & $-0.13$ & {$ 0.16$} & {$-0.03$} \\
    Scenario II  & {$1.58$} & {$4.04$} & {$0.49$} & $-0.13$ & {$ 0.19$} & {$ 0.01$} \\
    Scenario III & {$2.64$} & {$4.75$} & {$1.06$} & {$ 0.41$} & {$-0.18$} & {$ 0.54$} \\
    Scenario IV  & {$1.67$} & {$6.03$} & {$0.91$} & {$-0.22$} & {$ 1.18$} & {$ 2.01$} \\
    \bottomrule
  \end{tabular}
\end{table}

Scenario~III illustrates that a permanent 50\% reduction in cancer mortality across all ages substantially increases the products' standard deviation and reverses their usual skewness signs. The hedged portfolio partially absorbs this impact, but its standard deviation and skewness still rise to {$1.06$} and {$0.54$}, respectively. Scenario~IV's transitory but severe shocks concentrated on working-ages cause insurance standard deviation and skewness to significantly increase, while leaving the annuity only mildly affected. This imbalance weakens the underlying mechanism of natural hedging, and in turn causes portfolio standard deviation and skewness to increase substantially. Taken together, the four what-if scenarios demonstrate that the 3WPF-CLJ model accommodates a wide range of mortality dynamics, including positive and negative jumps, permanent, transitory and long-lasting effects, and age- and cause-specific heterogeneity, enabling comprehensive scenario analysis for insurance risk management.

\section{Conclusion}\label{sec:conclusion}

This paper presents a comprehensive framework for modeling age-specific mortality with cause-specific jump effects and long-lasting pandemic impacts. Our empirical analysis of U.S. mortality data reveals three critical features of the COVID-19 pandemic: substantial aggregate mortality shocks, heterogeneous impacts across age-cause combinations, and divergent persistence patterns across causes of death. These findings motivate the development of the 3WPF-CLJ model, which extends existing stochastic mortality frameworks by simultaneously capturing age-specific jump severity and cause-specific long-lasting effects through a flexible decay function.

The proposed model makes several methodological contributions. First, the three-way parallel factors structure incorporates both common mortality trends and cause-specific deviations, allowing the model to preserve distinct features across causes while capturing general mortality improvements. Second, the cause-specific lingering jump component uses a gamma-density-like decay function to explain the empirical heterogeneous persistence patterns. Third, the Route II estimation method efficiently handles the model's complexity by working with mortality improvement rates. Model comparisons suggest that both the jump component and cause-specific term are essential. Omitting jumps distorts estimates of long-term mortality trends, while omitting cause-specific heterogeneity leads to unrealistic averaging and hides important demographic patterns.

Our analysis further demonstrates the model's practicality in life insurance valuation and risk management. The natural hedging application shows that optimal portfolio construction requires accurate modeling of both age-specific and cause-specific mortality dynamics. Models that aggregate across causes or ages produce suboptimal hedge ratios, leading to substantially higher risk compared to the optimal hedge. The what-if scenarios demonstrate that the 3WPF-CLJ model can accommodate diverse mortality forecasts, including permanent mortality improvements from medical breakthroughs, endemic regimes with frequent mild shocks, and catastrophic events concentrated in specific age ranges. This flexibility enables life insurers to evaluate long-term liabilities and formulate risk management strategies under a wide range of plausible mortality scenarios.

{
Several limitations warrant acknowledgment. First, our dataset contains only one major pandemic event, which limits the precision of the jump probability estimate. Future research would benefit from incorporating longer sample periods covering multiple pandemics and wars to strengthen parameter inference and validate decay patterns across different types of mortality shocks.

Second, our empirical analysis focused on U.S. mortality data for a limited set of age groups and causes of death, reflecting the limited availability in other countries of comparable cause-of-death data that meet the requirements of this study. Extending the framework to individual ages, granular causes of death, other countries, and multi-population settings remains an important direction for future work.

Third, our risk management analysis examined only simplified insurance and annuity products under a basic natural hedging framework, and the results may differ for realistic portfolios with many products, ages and policy terms. Future investigations could apply the framework to index-based longevity hedging strategies or the pricing of standardized mortality securities.

Fourth, the persistence of the jump is estimated from only three post-pandemic years, and the decay is assumed to be common to all ages within a cause. As more post-pandemic data become available, the fitted decay can be reassessed and refined, for example by allowing it to differ across broad age groups.}

The COVID-19 pandemic has fundamentally challenged traditional assumptions about mortality risk in life insurance and annuity products. As the industry navigates post-pandemic uncertainty, stochastic mortality models that capture heterogeneous, long-lasting effects across ages and causes of death will be essential for robust risk assessment, portfolio construction, and reserve planning in the anticipation of future health crises.

\section*{Funding Acknowledgment}
Kenneth Q. Zhou acknowledges the support of the Natural Sciences and Engineering Research Council of Canada (RGPIN-2025-04157 and DGECR-2025-00488). 

\section*{Declaration of Generative AI Use}
During the preparation of this work the author(s) used ChatGPT and Claude in order to improve the language and readability of the paper. After using this tool/service, the author(s) reviewed and edited the content as needed and take(s) full responsibility for the content of the publication.

\section*{Declaration of Competing Interest}
The authors declare that they have no known competing financial interests or personal relationships that could have appeared to influence the work reported in this paper.

\section*{Data Availability}
Data used in this study are publicly available from the CDC WONDER system (\url{https://wonder.cdc.gov}).

\section*{CRediT Authorship}
\emph{Yanxin Liu:} Conceptualization, Data curation, Formal analysis, Investigation, Methodology, Validation, Visualization, Writing -- original draft, Writing -- review \& editing.
\emph{Kenneth Q.\ Zhou:} Conceptualization, Data curation, Formal analysis, Investigation, Methodology, Validation, Visualization, Writing -- original draft, Writing -- review \& editing.

\bibliographystyle{apalike}
\bibliography{Reference.bib}

\newpage
\appendix

\section{Notation} \label{app:notation}

{Table~\ref{tab:notation} summarizes the notation used in this paper for easy reference.}

\begin{table}[!htbp]
\centering

\caption{Notation used in this paper.}
\label{tab:notation}
\begin{tabularx}{\linewidth}{@{}
    >{\raggedright\arraybackslash}p{0.25\linewidth}
    >{\raggedright\arraybackslash}X
    @{}}
\toprule
Symbol & Meaning \\
\midrule

\multicolumn{2}{@{}l@{}}{
    \textit{Panel A: General notation and mortality measures}
} \\
\hdashline
\addlinespace
$x$, $t$, $c$
    & Age group, calendar year, and cause of death \\
$X$, $T$, $C$
    & Number of age groups, years, and causes \\
$m_{x,t,c}$
    & Central death rate at age $x$ in year $t$ from cause $c$ \\
$Z_{x,t,c}$
    & Log mortality improvement rate,
      $\ln m_{x,t,c} - \ln m_{x,t-1,c}$ \\
$\vec{\theta}$
    & Vector of all model parameters \\
$l(\vec{\theta})$
    & Log-likelihood function \\

\midrule
\multicolumn{2}{@{}l@{}}{
    \textit{Panel B: Baseline mortality dynamics}
} \\
\hdashline
\addlinespace
$a_{x,c}$
    & Level of log mortality at age $x$ for cause $c$ \\
$B_x$, $K_t$
    & Age pattern and period index of the factor common to all causes \\
$D$, $\sigma_\eta$
    & Drift and volatility of $K_t$; $\eta_t$ is its innovation \\
$\varphi_c$
    & Loading of cause $c$ on the cause-specific factor \\
$b_x$, $k_t$
    & Age pattern and period index of the cause-specific factor \\
$d$, $\sigma_\xi$
    & Drift and volatility of $k_t$; $\xi_t$ is its innovation \\
$e_{x,t,c}$, $\sigma_e$
    & Error term and its standard deviation \\

\midrule
\multicolumn{2}{@{}l@{}}{
    \textit{Panel C: Jump occurrence and magnitude}
} \\
\hdashline
\addlinespace
$N_t$
    & Indicator that a jump has occurred in or before year $t$ \\
$T_J$
    & Year in which the jump occurs \\
$p$
    & Probability that a jump occurs in a given year \\
$J_{x,c}$
    & Size of the initial jump at age $x$ for cause $c$ \\
$\mu_{x,c}$, $\sigma_J$
    & Mean and standard deviation of $J_{x,c}$ \\

\midrule
\multicolumn{2}{@{}l@{}}{
    \textit{Panel D: Jump persistence}
} \\
\hdashline
\addlinespace
$\tau$
    & Years elapsed since the jump, $\tau = t - T_J$ \\
$J_{x,c}^{(\tau)}$
    & Lingering jump effect $\tau$ years after the jump,
      $J_{x,c}\,\pi_c(\tau)$ \\
$\pi_c(\tau)$
    & Decay of the jump effect for cause $c$, with $\pi_c(0)=1$ \\
$\alpha_c$, $\beta_c$, $\gamma_c$
    & Shape, rate, and magnitude of the decay for cause $c$ \\

\bottomrule
\end{tabularx}
\end{table}
\clearpage
\section{Review of Existing Stochastic Models with Mortality Jumps and Their Insufficiency} \label{app:review}

In the literature, extreme mortality risk is typically modeled by stochastic mortality models with mortality jump effects. Many of the proposed mortality jump models are based on the Lee-Carter model \citep{LeeCarter1992} structure, which assumes 
\begin{equation}\label{eqn: LC model}
\ln(m_{x,t}) = a_x+b_xk_t+e_{x,t},
\end{equation}
where $a_{x}$ represents the static level of mortality rate at age $x$, $k_t$ captures the variation of log mortality  rates over time, and $b_x$ measures the sensitivity of log mortality rates to changes in $k_t$.

Some researchers model catastrophic mortality events as permanent mortality jumps, upon which systematic shifts are added to the long-term mortality dynamic. For example, \cite{Cox2006} proposed a permanent jump model where the jump effects are imposed into the evolution of mortality dynamics $k_t$ over time, that is
\begin{equation}
\label{eqn:Cox et al 2006}
k_t = \left\{ 
\begin{array}{ll}
d + k_{t-1} - p\times \text{E}(J_t) + \eta_t, & \text{if } N_{t+1}=0 \\
d + k_{t-1} - p\times \text{E}(J_t) + \eta_t + J_{t}, & \text{if } N_{t+1}=1 \\
\end{array}
\right.  
\end{equation}
where $d$ represents the drift of the stochastic process, $p$ is the probability of jump occurring in a year, $m$ is the expectation of the jump severity, $J_t$ is the jump variable and $\eta_t$ is the innovation term. This is a permanent jump model as a jump occurring in year $s$ would affect all future $k_t$ for $t\geq s$. 

To address the short-term severe effects of catastrophic mortality events (such as the 1918 Spanish Flu, and WWI/WWII), several researchers, including \cite{CoxSamuelH.2010MrmA} and \cite{ChenCox2009}, have considered transitory jumps whose effects vanish shortly after occurrence. For example, \cite{ChenCox2009} adapted equation \eqref{eqn:Cox et al 2006} to create a  transitory jump variant by separating the jump process from the general dynamic. More specifically, under the CC model we have
\begin{equation}
\label{eqn:ChenCox2009}
\left\{ 
\begin{array}{l}
\tilde{k}_t = d + \tilde{k}_{t-1} + \eta_t \\
k_{t} = \tilde{k}_t + N_tJ_t
\end{array}
\right.
\end{equation}
where $\tilde{k}_t$ captures the general evolution of mortality that is not affected by the jump effect. The jump $N_tJ_t$ only carries a transitory effect since it is added separately to $k_t$. Both $\eta_t$ and $J_t$ in the CC model are assumed to follow a Gaussian distribution, respectively. One limitation of this model is that it inherently assumes the jump effect and general mortality dynamic share the same age pattern. Such constraint is then relaxed by the J1/J2 model proposed by \cite{LiuLi2015}, which assumes
\begin{equation} \label{eqn: Liu and Li 2015}    
\ln(m_{x,t}) = a_x+b_xk_t+N_tJ_{x,t}+e_{x,t}
\end{equation}
In the above formulation, the jump variable $J_{x,t}$ is allowed to have varying age patterns that may or may not be the same as the general mortality trend. To be more specific, the jump effect vector $\vec{J}_t = (J_{1,t},\ldots,J_{X,t})^\top$ in the J1 model follows a multivariate Gaussian distribution, with mean and variance-covarince matrix satisfy
\[
\text{E}(\vec{J}_t) = \vec{\beta}^{(J)}\mu_J, \qquad \text{ and } \qquad \text{Var}(\vec{J}_{t})= \vec{\beta}^{(J)}(\vec{\beta}^{(J)})^\top\sigma^2_J,
\]
while model J2 further extend model J1 by capturing the correlations of jump effects across different age groups via an additional parameter $\rho$.

More recently, as a result of the outbreak of the COVID pandemic, several new mortality models have emerged to capture the most recent catastrophic mortality event and its impacts on human mortality. \cite{chen2022modeling} proposed a multi-population threshold jump mortality model, which contains a pandemic shock and a population-specific shock. According to their model, the pandemic jump occurs in a population if the pandemic event causes more deaths than the average new deaths of the world. \cite{VANBERKUM2025144} extend the multi-population model developed by \cite{LiLee2005} to a three-layer Li-Lee model. In their proposed model, the first two layers are identical to the original Li and Lee model which captures the populations' co-movement and the population-specific movement during the pre-COVID era. The last layer is specifically designed to capture the excess mortality from the pandemic after year 2019.

Focusing on the on-going effects of the pandemic,  
\cite{zhou2022multi} designed a multi-parameter-level mortality model which can be used for generating future catastrophic mortality scenarios. Their proposed model is based on the Lee-Carter structure, with the inclusion of a jump component. It is defined by
\[
\ln(m_{x,t}) = a_x+b_xk_t+c_{x,t}\pi_t \mathds{1}_{\{T\leq t \leq T+n\}}
\]
where $c_{x,t}$ and $\pi_t$ represent the age pattern and the size of the pandemic in year $t$, respectively. The indicator function $\mathds{1}_{\{T\leq t \leq T+n\}}$ imposes the expert opinion that the pandemic occurs in year $T$ and will last for $n$ years. According to \cite{zhou2022multi}, the proposed model uses three parameter levels which reflects (1) long-term mortality pattern, (2) COVID-related excess mortality effect, and (3) expert opinions regarding the likelihood of future pandemic occurrence. 

To address the long-lasting (or so-called ``vanishing'') effects of the COVID-19 pandemic, \cite{Goes2025} proposed a model that assumes an AR or MA structure in the jump process. Their proposed model is defined by
\[
\ln(m_{x,t}) = a_x+b_xk_t+b_x^{(J)}J_t+e_{x,t}
\]
where $J_t$ is the jump process that takes the form of
\[
\left\{
\begin{array}{rcll}
J_t & = & \phi_1 J_{t-1} + N_tY_t, &\quad\quad \text{if AR(1) is assumed, or}\\
J_t & = & N_tY_t + \theta_1 N_{t-1}Y_{t-1}, &\quad\quad \text{if MA(1) is assumed,}\\
\end{array}
\right.
\]
and $b_x^{(J)}$ captures the age pattern of the jump effects. In the jump process, $N_t$ is the indicator variable of jump occurrence, and $Y_t$ is the magnitude of the transitory jump effect. One key feature of the model developed by \cite{Goes2025} is the use of AR/MA structure in the jump process, which allows the model to capture jump effects that last over one year. However, for jumps that peak at later years, the model could over-estimate the occurrence probability of catastrophic mortality event if the effects of the pandemic does not vanish shortly after occurrence. 


\section{Estimation Procedure} \label{app:algorithm}

\subsection{Broyden-Fletcher-Goldfarb-Shanno (BFGS) Algorithm}
In this paper, the estimation procedure is carried out by a quasi-newton method called the Broyden-Fletcher-Goldfarb-Shanno (BFGS) algorithm \citep{Broyden1970,Fletcher1970,Goldfarb1970,Shanno1970}\footnote{Alternatively, parameter estimation can be carried out by the Expectation-Maximization (EM) algorithm, which is a commonly used method for the Gaussian mixture model. }. Compared to the traditional Newton method, the quasi-newton method uses the approximated Hessian matrix and is therefore useful when the exact Hessian is computationally expensive. We summarize the key steps of the BFGS algorithm in the iterative procedures described in Algorithm \ref{alg: BFGS}. 

\mbox{}\vspace{-\baselineskip}
\begin{tcolorbox}[box align=center, colback=white, colframe=black!75!white]
\captionof{algorithm}{BFGS Optimization Procedure}\label{alg: BFGS}
\rule{\linewidth}{0.4pt}
\begin{flushleft}
\textbf{Input:} Initial parameter vector $\vec{\theta}_0$; initial (approximate) Hessian matrix $\mathbf{H}_0$\\
\textbf{Parameter:} $\vec{\theta}$ includes $\vec{B}, \; D, \; \sigma_{\eta}, \; \vec{\varphi}, \; \vec{b}, \; d, \; \sigma_{\xi}, \; \vec{\mu}, \; \vec{\alpha}, \; \vec{\beta}, \; \vec{\gamma}, \; p, \; \sigma_{e}$\\
\textbf{Output:} Estimated parameter vector $\vec{\theta}$\\
\textbf{Note:} Gradients are approximated numerically using forward differences with a $1\%$ perturbation per parameter dimension.
\end{flushleft}
\vspace{0.2em}
\begin{algorithmic}[1]
\Procedure{BFGS Optimization}{}
    \State Evaluate initial log-likelihood: $l(\vec{\theta}_0)$ using equation \eqref{eqn: loglik}
    \State Set iteration counter $j \gets 0$
    \State Initialize convergence criterion: $R \gets 10$
    \While{$R > 10^{-8}$}
        \State Compute gradient (numerically) at current iterate:
        \[
        \nabla l(\vec{\theta}_j) = \left. \frac{\partial l}{\partial \vec{\theta}} \right|_{\vec{\theta} = \vec{\theta}_j}
        \]
        \State Solve for search direction: $\mathbf{H}_j \vec{s}_j = -\nabla l(\vec{\theta}_j)$
        \State Update parameters: $\vec{\theta}_{j+1} \gets \vec{\theta}_j + \vec{s}_j$
        \State Evaluate updated log-likelihood: $l(\vec{\theta}_{j+1})$
        \State Compute relative change:
        \[
        R = \frac{l(\vec{\theta}_{j+1}) - l(\vec{\theta}_j)}{l(\vec{\theta}_j)}
        \]
        \State Compute gradient difference: $\vec{y}_j \gets \nabla l(\vec{\theta}_{j+1}) - \nabla l(\vec{\theta}_j)$
        \State Update inverse Hessian approximation:
        \[
        \Delta \mathbf{H}_j = \frac{\vec{y}_j \vec{y}_j^\top}{\vec{y}_j^\top \vec{s}_j} - \frac{\mathbf{H}_j \vec{s}_j \vec{s}_j^\top \mathbf{H}_j}{\vec{s}_j^\top \mathbf{H}_j \vec{s}_j}
        \]
        \[
        \mathbf{H}_{j+1} \gets \mathbf{H}_j + \Delta \mathbf{H}_j
        \]
        \State Increment iteration: $j \gets j + 1$
    \EndWhile
    \State \Return{$\vec{\theta}_{j-1}$}
\EndProcedure
\end{algorithmic}
\end{tcolorbox}

\subsection{Initial Values}

To implement the BFGS method, we are required to specify the initial value of the parameter vector $\vec{\theta}$, which is crucial to ensure fast convergence of the estimation procedure. In this paper, we determine the initial values for each of the parameters in $\vec{\theta}$ using the following procedure.
\begin{itemize}
    \item For parameters related to the general mortality trend $\vec{B}$, $D$, and $\sigma_\eta$: \\We fit the aggregate mortality rate to the traditional Lee-Carter model,
    \[
    \ln(m_{x,t}) = a_x + B_xK_t
    \]
    using maximum likelihood estimation where the aggregate death count is assumed to follow a Poisson distribution, $D_{x,t}\sim Poisson(E_{x,t}m_{x,t})$ \citep{BROUHNS2002373}. To exclude the COVID effect, only data prior to 2020 is used. The initial value of $\vec{B}$ is set to be the estimates of $\vec{B}$. The initial values of $D$ and $\sigma_\eta$ are set to be the mean and estimated standard deviation of the first difference of $K_t$. 
    \item For parameters related to the cause-specific mortality trend $\vec{\varphi}$, $\vec{b}$, $d$, and $\sigma_\xi$: \\ We consider the cause-specific mortality rates $\ln(m_{x,t,c})$ and compute their averages over time, $a_{x,c} = \sum_{t=1}^T\ln(m_{x,t,c})/T$. Then we back out the cause-specific residuals by $e_{x,t,c} = \ln(m_{x,t,c}) -  a_{x,c} - B_xK_t$, where $B_x$ and $K_t$ are the Lee-Carter estimates using the aggregate mortality rates. Finally, we fit the cause-specific residuals $e_{x,t,c}$ to a 3-way PARAFAC model to retrieve the initial values of $\vec{\varphi}$, $\vec{b}$, and $\vec{k}$. The initial values for the drift and the standard deviation of $k_t$ are set to be the mean and estimated standard deviation of the first difference of $k_t$. Similarly, only data prior to 2020 is used to exclude the COVID effect.

    \item For parameter related to the noise $\sigma_e$: \\
    We remove both general and cause-specific mortality trend from the cause-specific mortality rate $e^\ast_{x,t,c} = \ln(m_{x,t,c}) -  a_{x,c} - B_xK_t - \varphi_cb_xk_t$. The initial value of $\sigma_e$ is set to be the estimated standard deviation of $e^\ast_{x,t,c}$.
    
    \item For parameters related to the jump size $\vec{\mu}$: \\
    We consider the (log) mortality improvement rates in year 2020, $Z_{x,2020,c} = \ln(m_{x,2020,c}) - \ln(m_{x,2019,c})$. The size of the jump effect is initialized by $\mu_{x,c} = Z_{x,2020,c} - B_x\times D - \varphi_c\times b_x\times d$, which removes the one-year regular improvement. {The jump severity volatility $\sigma_J$ is set to zero and is not estimated, since we only have one jump observation in the data.}

    \item For parameters related to the long-lasting effects $\vec{\alpha}, \vec{\beta},$ and $\vec{\gamma}$: \\ Let us consider the cause-specific (log) mortality improvement rates after year 2020. We define 
    \[
    Z^\ast_{x,t,c} = Z_{x,t,c} - B_x\times (t-2019)D - \varphi_c\times b_x\times (t-2019)d
    \] for year 2020 to 2023, which excludes the effects from general and cause-specific mortality improvement. We then initialize the long-lasting effect parameters via the minimization of sum of squares residuals post COVID period, that is,
    \[
    \sum_{t=2021}^{2023} \left( Z^\ast_{x,t,c} - \mu_{x,c}\times\pi_c(t-2020) \right)^2
    \]
    where $\pi_c(t-2020)$ follows equation \eqref{eqn: def_LLE}.
    \item For parameter related to the jump frequency $p$: \\ The initial value of the jump occurrence is set to be 1/$T$ where $T=56$ according to the data set used in this paper.
 
\end{itemize}

\subsection{Potential Issues: Data with High Dimension}

The proposed 3WPF-CLJ model can be applied to data set with individual ages or more subdivided causes of death. However, when considering high-dimensional data, the estimation of the proposed model may be computationally expensive. To address this potential issue, we may consider a special case of the proposed 3WPF-CLJ model, under which all underlying gradients can be solved analytically, making the BFGS algorithm easier to converge. The estimated values of this special model can be treated as the initial values for the proposed 3WPF-CLJ model. 

To be more specific, we consider the special case of the proposed 3WPF-CLJ model with the following two simplifying assumptions: (1) the jump severity is non-random; and (2) there is no long-lasting effect of mortality jump. By letting 
\begin{equation}
    \sigma_J = 0 \quad \text{ and } \quad \pi_c(\tau)= \left\{ 
\begin{array}{ll}
1  & \text{ when } \tau = 0 \\
0  & \text{ when } \tau \neq 0 \\
\end{array} \right.,
\end{equation}
we can rewrite $f\left(\vec{z} \; | \; \vec{N} \; ; \;\vec{\theta} \;\right)$ into a product of conditional probabilities,
\begin{equation}
\resizebox{\textwidth}{!}{    
$\begin{array}{rcl}
\displaystyle f\left(\vec{z} \; |\vec{N};\vec{\theta}\right) & = &  \displaystyle f\left(\vec{z}_T\left|\vec{z}_{2},\ldots,\vec{z}_{T-1}\right.,\vec{N}; \;\vec{\theta} \;\right) \times f\left(\vec{z}_{T-1}\left|\vec{z}_{2},\ldots,\vec{z}_{T-2}\right., \vec{N}; \;\vec{\theta} \;\right)\times \cdots   \times f\left(\vec{z}_{3}\left|\vec{z}_2\right., \vec{N}; \;\vec{\theta} \;\right) \times f\left(\vec{z}_{2} \;| \vec{N} ; \;\vec{\theta} \;\right) \\
 & = & \displaystyle f\left(\vec{z}_T\left|\vec{z}_{T-1} \right., \vec{N} ; \;\vec{\theta} \;\right) \times f\left(\vec{z}_{T-1}\left|\vec{z}_{T-2}\right., \vec{N}; \;\vec{\theta} \;\right)\times \cdots \times f\left(\vec{z}_{3}\left|\vec{z}_2\right., \vec{N}; \;\vec{\theta} \;\right) \times f\left(\vec{z}_{2} \;| \vec{N} ; \;\vec{\theta} \;\right) \\
& = & \displaystyle \frac{ \prod_{t=2}^{T-1}\left(  f\left( \vec{z}_{t}, \vec{z}_{t+1} \;| \vec{N} ; \;\vec{\theta} \; \right)  \right)}{ \prod_{t=3}^{T-1}\left( f\left( \vec{z}_{t} \; | \vec{N} ; \;\vec{\theta} \; \right)  \right)}  
\label{eqn: conditional density simp}
\end{array}$}
\end{equation}
using the fact that $\vec{Z}_t$ is uncorrelated with $\vec{Z}_s$ for $s<t-1$ when the jump severity is non-random and the time lag is greater than 1. Under simplifying assumptions, $f(\vec{z} \; |\vec{N};\vec{\theta})$ comprises marginal density $f(\vec{z}_t\; ; \;\vec{\theta} \; )$ with $(X \times C)$ dimensions and joint density $f(\vec{z}_t, \vec{z}_{t+1}\; ; \;\vec{\theta} \; )$ with $2(X\times C)$ dimensions. The dimension required to evaluate the conditional probability $f(\vec{z} \; |\vec{N};\vec{\theta})$ has reduced significantly. The log-likelihood function of this special case can be evaluated using equations \eqref{eqn: loglik} to \eqref{eqn: complete joint density v2} with conditional density $f(\vec{z} \; |\vec{N};\vec{\theta})$ replaced by equation \eqref{eqn: conditional density simp}. 

\noindent{\bf Conditional Marginal Density}\\
Let us use the following simplified notation to denote the conditional marginal density:
\[
f_t^{n_{t-1},n_{t}} := f(\vec{z}_t|N_{t-1}=n_{t-1}, N_{t}=n_{t}\; ; \;\vec{\theta} \; )
\]
Conditional on $N_{t-1}$ and $N_t$, $\vec{Z}_t$ follows a multivariate normal distribution,
\[
\vec{Z}_t|N_{t-1}=n_{t-1},N_t=n_{t} \sim \text{MVN}(\vec{\zeta}_{n_{t-1},n_{t}}, {\bf\Sigma}_{n_{t-1},n_{t}})
\]
where both the mean vector $\vec{\zeta}_{n_{t-1},n_{t}}$ and the variance-covariance matrix ${\bf\Sigma}_{n_{t-1},n_{t}}$ are functions of the realization of $N_{t-1}=n_{t-1}$ and $N_t=n_{t}$,  
\[
\vec{\zeta}_{n_{t-1},n_{t}} = \vec{1}_{C} \otimes \vec{B}\times D+\vec{\varphi}\otimes\vec{b}\times d+\left(n_{t}-n_{t-1}\right)\times\vec{\mu}^{(J)}
\]
and 
\[
{\bf\Sigma}_{n_{t-1},n_{t}}    =  \left(\vec{1}_{C} \otimes \vec{B}\right)\left(\vec{1}_{C} \otimes \vec{B}\right)^\top \times\sigma_{\eta}^2  + \left(\vec{\varphi}\otimes\vec{b}\right)\left(\vec{\varphi}\otimes\vec{b}\right)^\top \times\sigma_{\xi}^2 + 2\times\mathbf{I}_{XC}\times \sigma_e^2,
\]
The density of $\vec{Z}_t|N_{t-1},N_t$ can then be written down explicitly as
\[
f_t^{n_{t-1},n_{t}} = \text{det}(2\pi{\bf\Sigma}_{n_{t-1},n_{t}})\text{exp}\left( -\frac{1}{2} \left(\vec{z}_t-\vec{\zeta}_{n_{t-1},n_{t}}\right)^\top\left({\bf\Sigma}_{n_{t-1},n_{t}}\right)^{-1}\left(\vec{z}_t-\vec{\zeta}_{n_{t-1},n_{t}}\right) \right)
\]

\noindent{\bf Conditional Joint Density}\\
Similarly, let us use the following notation to denote the conditional joint density:
\[
f_{t,t+1}^{n_{t-1},n_{t},n_{t+1}} := f(\vec{z}_t, \vec{z}_{t+1}|N_{t-1}=n_{t-1}, N_{t}=n_{t},N_{t+1}=n_{t+1}\; ; \;\vec{\theta} \; )
\]
Conditional on $N_{t-1}$, $N_t$ and $N_{t+1}$, the vector of $\vec{Z}^\ast_{t,t+1} = (\vec{Z}_t, \vec{Z}_{t+1})$ also follows a multivariate normal distribution, 
\[
\left. \vec{Z}_{t,t+1}^\ast \right| N_{t-1}=n_{t-1},N_t=n_{t}, N_{t+1}=n_{t+1} \sim \text{MVN}(\vec{\zeta}^\ast_{n_{t-1},n_{t},n_{t+1}}, {\bf\Sigma}^\ast_{n_{t-1},n_{t},n_{t+1}})
\]
where
\[
\vec{\zeta}_{n_{t-1},n_{t},n_{t+1}}^\ast = \left( \begin{matrix} \vec{\zeta}_{n_{t-1},n_{t}}  \\ \vec{\zeta}_{n_{t},n_{t+1}} \end{matrix}\right)
\]
and
\[
{\bf\Sigma}^\ast_{n_{t-1},n_{t},n_{t+1}} = \left( \begin{matrix} {\bf\Sigma}_{n_{t-1},n_{t}} & {\bf O}\\ {\bf O} & {\bf\Sigma}_{n_{t},n_{t+1}}\end{matrix}\right),
\]
with ${\bf O} =  - \mathbf{I}_{XC} \times \sigma_e^2$ representing the off-diagonal term.
The density of $\vec{Z}^\ast_{t,t+1}$ conditional on $N_{t-1}=n_{t-1},N_t=n_{t},N_{t+1}=n_{t+1}$ can then be written down explicitly as
\[
f_{t,t+1}^{n_{t-1},n_{t},n_{t+1}} = \text{det}(2\pi{\bf\Sigma}^\ast_{n_{t-1},n_{t},n_{t+1}})\text{exp}\left( -\frac{1}{2} \left(\vec{z}^\ast_{t,t+1}-\vec{\zeta}^\ast_{n_{t-1},n_{t},n_{t+1}}\right)^\top\left({\bf\Sigma}^\ast_{n_{t-1},n_{t},n_{t+1}}\right)^{-1}\left(\vec{z}^\ast_{t,t+1}-\vec{\zeta}^\ast_{n_{t-1},n_{t},n_{t+1}}\right) \right)
\]

\noindent{\bf Derivative Computation}\\
Recall that we use $\vec{\theta}$ to represent the vector of all the parameters in the model, and $\theta_0$ to represent a particular parameter in $\vec{\theta}$. Since that the log-likelihood function is given by
\[    
l(\vec{\theta}) = \ln\left( f\left(\vec{z} \, ; \,\vec{\theta} \right) \right)
\]
where
\[
    \begin{array}{cclc}
\displaystyle f\left(\vec{z} \, ; \,\vec{\theta} \right) & = & f\left(\vec{z} \, | \, N_1=\cdots=N_T=0 \, ; \,\vec{\theta} \right)\times (1-p)^T & \fbox{\text{No Jump}}\\[0.2cm]
\displaystyle     &  & + \, f\left(\vec{z} \, | \, N_1=\cdots=N_{T-1}=0, N_{T}=1 \, ; \,\vec{\theta} \right)\times (1-p)^{T-1}\times p & \fbox{\text{Jump in Year $T$}}\\[0.2cm]
\displaystyle     &  & + \, \cdots & \vdots\\[0.2cm]
\displaystyle     &  & + \, f\left(\vec{z} \, | \, N_1=\cdots=N_T=1 \, ; \,\vec{\theta} \right)\times p, & \fbox{\text{Jump in Year 1}}
\end{array} 
\]
The partial derivative of $l(\vec{\theta})$ with respect to $\theta_0$ can be computed by 
\[
\frac{\partial l(\vec{\theta})}{\partial \theta_0} = \left. \frac{\partial  f(\vec{z} \, ; \,\vec{\theta} )}{\partial \theta_0} \middle/ f(\vec{z} \, ; \,\vec{\theta} ) \right.
\]
where $\frac{\partial  f(\vec{z} \, ; \,\vec{\theta} )}{\partial \theta_0}$ follows one of the following two cases.\\

\noindent {\bf Case 1: when $\theta_0 = p$.} \\
\[
\begin{array}{ccl}
\displaystyle \frac{\partial  f(\vec{z} \, ; \,\vec{\theta} )}{\partial \theta_0} & = & -f\left(\vec{z} \, | \, N_1=\cdots=N_T=0 \, ; \,\vec{\theta} \right)\times T \times (1-p)^{T-1} \\[0.2cm]
\displaystyle & & + \sum\limits_{i=1}^T \left( (1-p)^{T-i} - (T-i)(1-p)^{T-i-1}p \right) f\left(\vec{z}|N_1=\cdots=N_{T-i}=0,N_{T-i+1}=\cdots=N_{T}=1;\vec{\theta}\right)
\end{array} 
\]

\noindent {\bf Case 2: when $\theta_0 \neq p$.} \\
\[
    \begin{array}{ccl}
\displaystyle \frac{\partial  f(\vec{z} \, ; \,\vec{\theta} )}{\partial \theta_0} & = &  \frac{\partial  }{\partial \theta_0}f\left(\vec{z} \, | \, N_1=\cdots=N_T=0 \, ; \,\vec{\theta} \right)\times (1-p)^T \\[0.2cm]
\displaystyle     &  & + \, \frac{\partial  }{\partial \theta_0}f\left(\vec{z} \, | \, N_1=\cdots=N_{T-1}=0, N_{T}=1 \, ; \,\vec{\theta} \right)\times (1-p)^{T-1}\times p \\[0.2cm]
\displaystyle     &  & + \, \cdots \\[0.2cm]
\displaystyle     &  & + \, \frac{\partial  }{\partial \theta_0}f\left(\vec{z} \, | \, N_1=\cdots=N_T=1 \, ; \,\vec{\theta} \right)\times p 
\end{array} 
\]
In the above formula, the partial derivatives of the conditional density can be computed by
\begin{equation}
\begin{array}{rcl}
\displaystyle \frac{\partial  }{\partial \theta_0}f\left(\vec{z} \; |\vec{N};\vec{\theta}\right) & = & \displaystyle \frac{ \prod_{t=2}^{T-1}\left(  f\left( \vec{z}_{t}, \vec{z}_{t+1} \;| \vec{N} ; \;\vec{\theta} \; \right)  \right)}{ \prod_{t=3}^{T-1}\left( f\left( \vec{z}_{t} \; | \vec{N} ; \;\vec{\theta} \; \right)  \right)}  \times \left( \Psi_{joint} - \Psi_{marginal}  \right)
\end{array}
\end{equation}
where

\[
\begin{array}{ll}
\displaystyle\Psi_{marginal} = &\displaystyle \sum\limits_{t=3}^{T-1} \left[ -\frac{1}{2}\text{Tr}\left( ({\bf\Sigma_{n_{t-1},n_{t}}})^{-1} \frac{\partial {\bf\Sigma_{n_{t-1},n_{t}}}}{\partial \theta_0} \right) + (\frac{\partial \vec{\zeta}_{n_{t-1},n_{t}}}{\partial \theta_0})^\top ({\bf\Sigma_{n_{t-1},n_{t}}})^{-1} (\vec{z}_t-\vec{\zeta}_{n_{t-1},n_{t}}) \right. \\
     &  \displaystyle \left. \qquad + \frac{1}{2} (\vec{z}_t-\vec{\zeta}_{n_{t-1},n_{t}})^\top  ({\bf\Sigma_{n_{t-1},n_{t}}})^{-1} \frac{\partial {\bf\Sigma_{n_{t-1},n_{t}}}}{\partial \theta_0}   ({\bf\Sigma_{n_{t-1},n_{t}}})^{-1}  (\vec{z}_t-\vec{\zeta}_{n_{t-1},n_{t}}) \right]
\end{array}
\]
and
\[
\begin{array}{ll}
\displaystyle  \Psi_{joint} =  &\displaystyle \sum\limits_{t=2}^{T-1} \left[ -\frac{1}{2}\text{Tr}\left( ({\bf\Sigma_{n_{t-1},n_{t},n_{t+1}}^\ast})^{-1} \frac{\partial {\bf\Sigma_{n_{t-1},n_{t},n_{t+1}}^\ast}}{\partial \theta_0} \right) + (\frac{\partial \vec{\zeta}_{n_{t-1},n_{t},n_{t+1}}^\ast}{\partial \theta_0})^\top ({\bf\Sigma_{n_{t-1},n_{t},n_{t+1}}^\ast})^{-1} (\vec{z}_{t,t+1}-\vec{\zeta}_{n_{t-1},n_{t},n_{t+1}}^\ast) \right. \\
     &  \displaystyle \left. \qquad + \frac{1}{2} (\vec{z}_{t,t+1}-\vec{\zeta}_{n_{t-1},n_{t},n_{t+1}}^\ast)^\top  ({\bf\Sigma_{n_{t-1},n_{t},n_{t+1}}^\ast})^{-1} \frac{\partial {\bf\Sigma_{n_{t-1},n_{t},n_{t+1}}^\ast}}{\partial \theta_0}   ({\bf\Sigma_{n_{t-1},n_{t},n_{t+1}}^\ast})^{-1}  (\vec{z}_{t,t+1}-\vec{\zeta}_{n_{t-1},n_{t},n_{t+1}}^\ast) \right]
\end{array}
\]
with 
\[
\frac{\partial \vec{\zeta}_{n_{t-1},n_{t},n_{t+1}}^\ast}{\partial \theta_0} = \left( \begin{matrix}
   \displaystyle \frac{\partial \vec{\zeta}_{n_{t-1},n_{t}}}{\partial \theta_0} \\
   \displaystyle \frac{\partial \vec{\zeta}_{n_{t},n_{t+1}}}{\partial \theta_0}
\end{matrix}\right) \quad \text{and} \quad \frac{\partial {\bf\Sigma_{n_{t-1},n_{t},n_{t+1}}^\ast}}{\partial \theta_0} = \left( \begin{matrix} \displaystyle\frac{\partial {\bf\Sigma_{n_{t-1},n_{t}}}}{\partial \theta_0} & \displaystyle \frac{\partial {\bf O}}{\partial \theta_0} \\ \displaystyle \frac{\partial {\bf O}}{\partial \theta_0}  & \displaystyle \frac{\partial {\bf\Sigma_{n_{t},n_{t+1}}}}{\partial \theta_0} \end{matrix} \right)
\]
The formulas of the generic $\frac{\partial \vec{\zeta}_{i,j}}{\partial \theta_0}$ and $\frac{\partial {\bf\Sigma_{i,j}}}{\partial \theta_0}$ are summarized in Table \ref{tab: pd_mean} and \ref{tab: pd_var}. For $\frac{\partial {\bf O_{n_{t}}}}{\partial \theta_0}$, we have
{\footnotesize\[
\frac{\partial {\bf O}}{\partial \theta_0} = \left\{ 
\begin{array}{ll}
- 2\times \mathbf{I}_{XC} \times \sigma_e     & \text{if} \quad \theta_0 = \sigma_e \\
{\bf 0} & \text{otherwise}
\end{array}
\right.
\]}

\begin{table}[!htbp]
\small
    \centering
    \caption{Formulas for the partial derivative of $\vec{\zeta}_{i,j}$ with respect to $\theta_0$ when $\theta_0 \neq p$. }
    \begin{tabular}{c|c}
       \hline
       \hline
       Parameter ($\theta_0$)  & Partial Derivative Formula\\
       \hline
       \multicolumn{2}{l}{- Related to general trend} \\
       \hline
       $B_x$  & $\vec{1}_C \otimes \vec{1}^\ast_x \times D$  \\
       $D$ &  $\vec{1}_C \otimes \vec{B}$\\
       $\sigma_\eta$ & $\mathbf{0}$ \\
       \hline
       \multicolumn{2}{l}{- Related to cause-specific trend} \\
       \hline
       $\varphi_c$  & $\vec{1}_c^\ast \otimes \vec{b} \times d$ \\
       $b_x$  & $\vec{\varphi} \otimes \vec{1}^\ast_x \times d$ \\
       $d$ &  $\vec{\varphi} \otimes \vec{b}$\\
       $\sigma_\xi$ & $\mathbf{0}$ \\
       \hline
       \multicolumn{2}{l}{- Related to jump effects} \\
       \hline
       $\mu_{x,c}$  & $(j-i) \times \vec{1}^\ast_{x+(c-1)C}$ \\
       \hline
       \multicolumn{2}{l}{- Related to errors} \\
       \hline
       $\sigma_e$  & $\mathbf{0}$ \\
       \hline
       \hline
       \multicolumn{2}{l}{\footnotesize Note: $\vec{1}^\ast_x = (0,\ldots,0, 1, 0, \ldots, 0)^\top$ is a vector of 0's with 1 in the $x$-th row. Similar definition applies to $\vec{1}^\ast_c$.}
    \end{tabular}
    \label{tab: pd_mean}
\end{table}

\begin{table}[!htbp]
\small
\centering
    \caption{Formulas for the partial derivative of ${\bf\Sigma_{i,j}}$ with respect to $\theta_0$ when $\theta_0 \neq p$. }
    \begin{tabular}{c|c}
       \hline
       \hline
       Parameter ($\theta_0$)  & Partial Derivative Formula\\
       \hline
       \multicolumn{2}{l}{- Related to general trend} \\
       \hline
       $B_x$  & $(\vec{1}_C\vec{1}_C^\top) \otimes (\vec{1}^\ast_x\vec{B}^\top+\vec{B}(\vec{1}^\ast_x)^\top) \times \sigma^2_\eta$  \\
       $D$ &  $\mathbf{0}$\\
       $\sigma_\eta$ & $\left(\vec{1}_{C} \otimes \vec{B}\right)\left(\vec{1}_{C} \otimes \vec{B}\right)^\top \times 2\sigma_\eta$ \\
       \hline
       \multicolumn{2}{l}{- Related to cause-specific trend} \\
       \hline
       $\varphi_c$  & $(\vec{1}^\ast_c\vec{\varphi}^\top+\vec{\varphi}(\vec{1}^\ast_c)^\top) \otimes (\vec{b}\vec{b}^\top)  \times \sigma^2_\xi$ \\
       $b_x$  & $(\vec{\varphi}\vec{\varphi}^\top) \otimes (\vec{1}^\ast_x\vec{b}^\top+\vec{b}(\vec{1}^\ast_x)^\top) \times \sigma^2_\xi$ \\
       $d$ &  $\mathbf{0}$\\
       $\sigma_\xi$ & $\left(\vec{\varphi}\otimes\vec{b}\right)\left(\vec{\varphi}\otimes\vec{b}\right)^\top \times 2\sigma_\xi$ \\
       \hline
       \multicolumn{2}{l}{- Related to jump effects} \\
       \hline
       $\mu_{x,c}$  & $\mathbf{0}$ \\
       \hline
       \multicolumn{2}{l}{- Related to errors} \\
       \hline
       $\sigma_e$  & $4\times\mathbf{I}_{XC} \times \sigma_e$ \\
       \hline
       \hline
       \multicolumn{2}{l}{\footnotesize Note: $\vec{1}^\ast_x = (0,\ldots,0, 1, 0, \ldots, 0)^\top$ is a vector of 0's with 1 in the $x$-th row. Similar definition applies to $\vec{1}^\ast_c$.}
    \end{tabular}
    \label{tab: pd_var}
\end{table}

\newpage
\section{Estimation Results for the U.S. Female Population} \label{app: U.S. female result}

\begin{figure}[ht!]
    \centering
    \includegraphics[width=\textwidth]{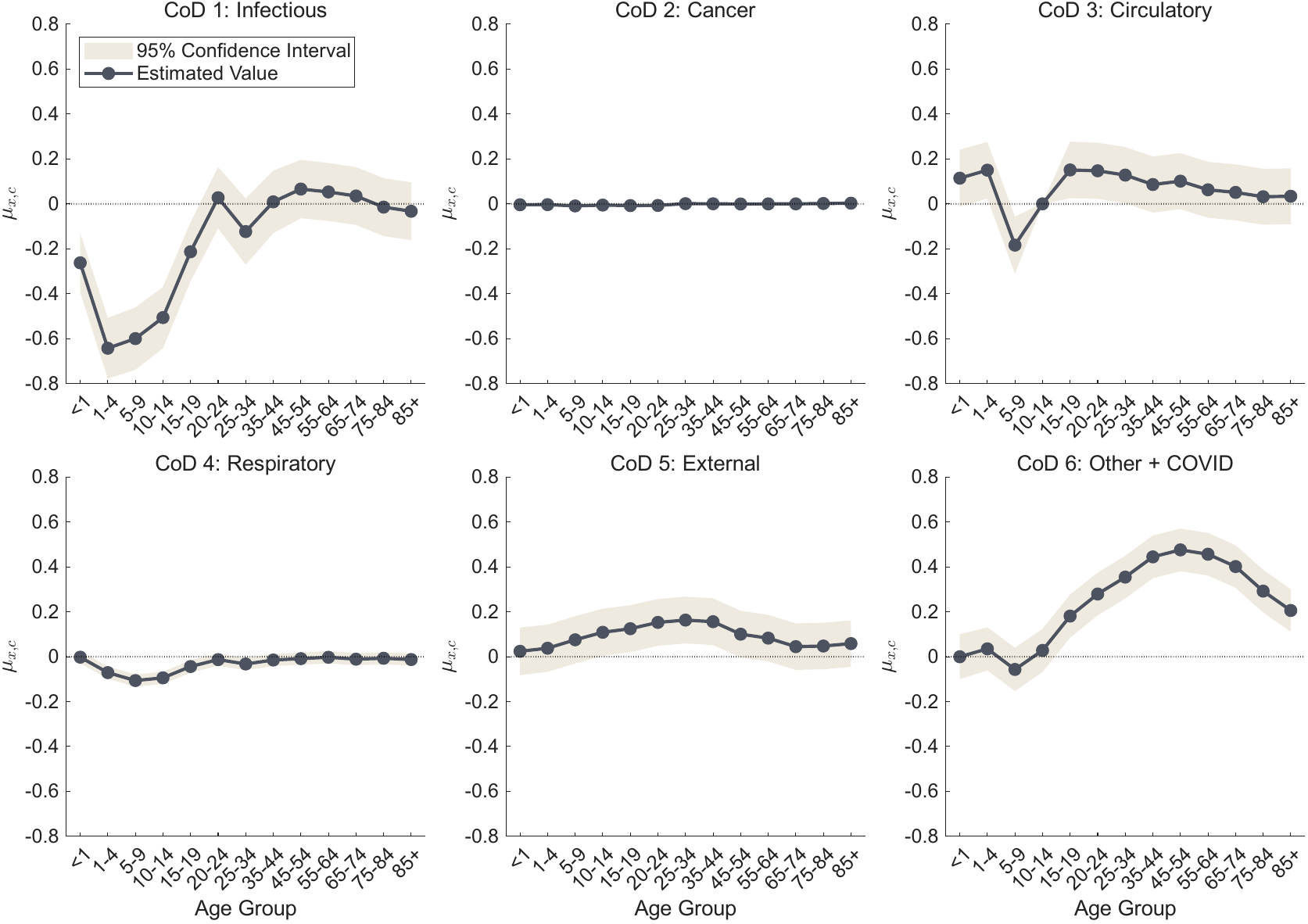}

    \caption{ Estimated values of $\mu_{x,c}$ with 95\% confidence intervals from the 3WPF-CLJ model fitted to U.S. female mortality data from 1968 to 2023.}
    \label{fig:JumpMu - female}
\end{figure}

{
In this appendix, we present the estimation results for the 3WPF-CLJ model fitted to U.S. female mortality data from 1968 to 2023, complementing the male results in Section \ref{sec:MdlAnl_Est}. Figure \ref{fig:Bxbx - female} reports the estimated common age pattern $B_x$ and the age profile $b_x$ of the cause-specific component, along with their 95\% confidence intervals. As in the male population, $B_x$ generally decreases with age. Combined with the negative estimated drift of the common period factor, this pattern indicates that older individuals experience smaller mortality improvements than younger individuals.

For cause-specific age deviations $b_x$, it resembles a mixture of two normal densities with peaks at ages 5-9 and 25–34, respectively. Compared with the male estimates, the first peak occurs one age group earlier and is more pronounced, while the second occurs at the same age group but with smaller magnitude. The contribution of this age profile to each cause is scaled by $\varphi_c$. According to the model specification, the loading for CoD 1 (infectious diseases) is fixed at one for model identification, while CoD 4 (respiratory diseases) captures the second largest loading with $\varphi_4 \approx 0.2083$.

\begin{figure}[!ht]
    \centering
    \includegraphics[width=\textwidth]{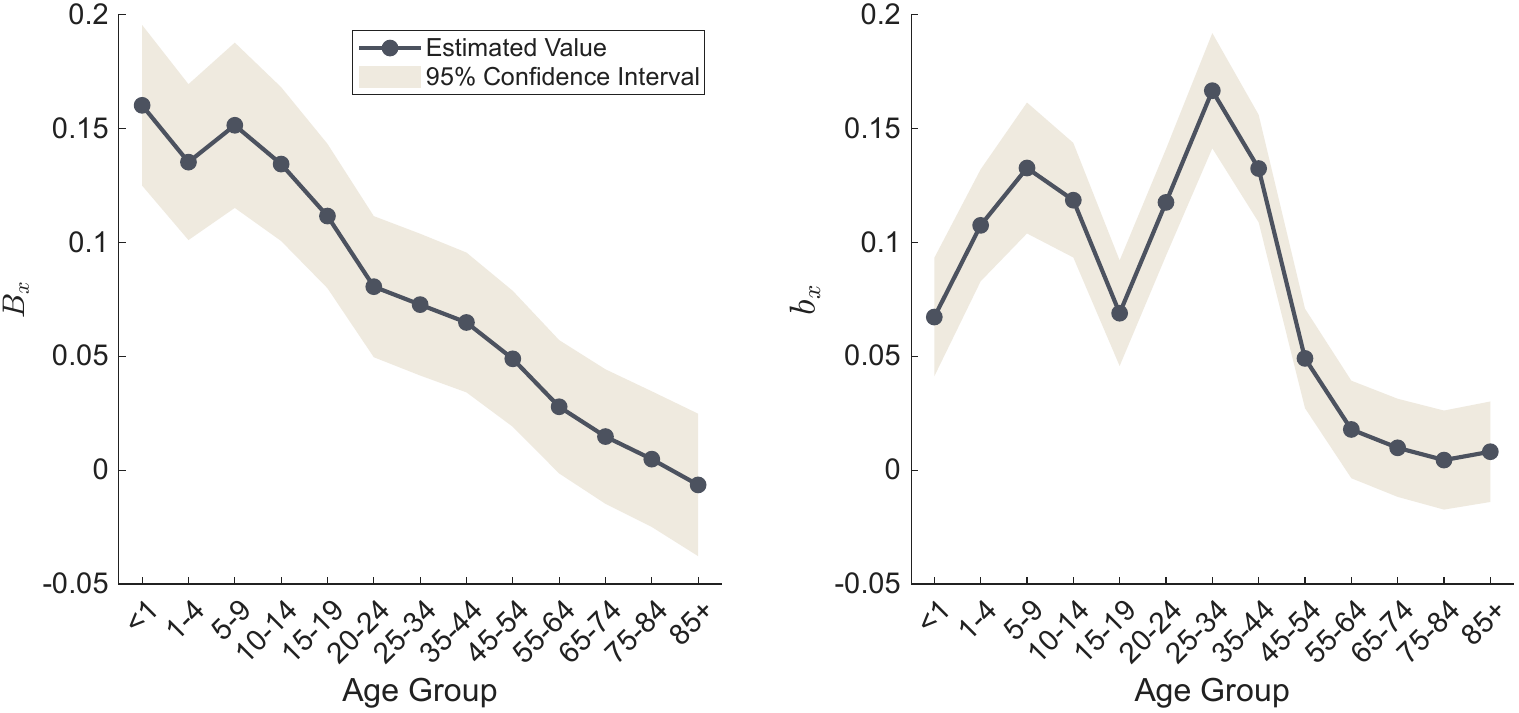}

    \caption{Estimated values of $B_x$ and $b_x$ with 95\% confidence intervals from the 3WPF-CLJ model fitted to U.S. female mortality data from 1968 to 2023.}
    \label{fig:Bxbx - female}
\end{figure}

Figure \ref{fig:JumpMu - female} presents the estimated jump severity levels $\mu_{x,c}$ and their 95\% confidence intervals for the six causes of death. The general shapes of the COVID-19 jump effects on the six causes resemble those estimates for males, although their magnitudes differ. For CoD 1 (infectious diseases), more negative effects are observed during childhood of females. The lowest estimate of the jump effects occurs at ages 1-4. Additionally, females generally have smaller jump severities at working ages for CoD 5 (external causes) and CoD 6 (others and COVID). Table \ref{tab:EstOthers - female} summarizes the estimated values and standard errors of the remaining parameters in the 3WPF-CLJ model for female population. As with the male results, the estimated jump effects characterize the observed pandemic event under the assumed model specification, while their longer-term persistence is inferred from a limited post-jump observation period from 2021-2023.
}

\begin{table}[!ht]
\centering

  \caption{ Parameter estimates with standard errors (S.E.) from the 3WPF-CLJ model (U.S. female mortality, 1968--2023).}
  \label{tab:EstOthers - female}
  \begin{tabular}{@{}lrr@{\hspace{1em}}lrr@{\hspace{1em}}lrr@{}}
    \toprule
    \multicolumn{9}{c}{\normalsize{General Parameters}} \\
    \cmidrule(lr){1-3} \cmidrule(lr){4-6} \cmidrule(lr){7-9}
    Parameter & \multicolumn{1}{c}{Estimate} & \multicolumn{1}{c}{S.E.} & Parameter & \multicolumn{1}{c}{Estimate} & \multicolumn{1}{c}{S.E.} & Parameter & \multicolumn{1}{c}{Estimate} & \multicolumn{1}{c}{S.E.} \\
    \midrule
$\varphi_1$ & $1.0000^{\dagger}$ & -- &
$D$ & $-0.1540$ & $0.0300$ &
$\log(\sigma_\eta)$ & $-1.5289$ & $0.1285$ \\
$\varphi_2$ & $0.0027$ & $0.0353$ &
$d$ & $0.0243$ & $0.1223$ &
$\log(\sigma_\xi)$ & $-0.1032$ & $0.1040$ \\
$\varphi_3$ & $0.0755$ & $0.0359$ &
$\mathrm{logit}(p)$ & $-3.9532$ & $1.0086$ &
$\log(\sigma_J)$ & $-\infty^{\dagger}$ & -- \\
$\varphi_4$ & $0.2083$ & $0.0463$ &
 & & &
$\log(\sigma_e)$ & $-2.6582$ & $0.0109$ \\
$\varphi_5$ & $0.0471$ & $0.0363$ & & & & & & \\
$\varphi_6$ & $0.0437$ & $0.0358$ & & & & & & \\
    \midrule
    \multicolumn{9}{c}{\normalsize{Long-Lasting Effect Parameters}} \\
    \cmidrule(lr){1-3} \cmidrule(lr){4-6} \cmidrule(lr){7-9}
$\log(\alpha_c)$ & Estimate & S.E. &
$\log(\beta_c)$ & Estimate & S.E. &
$\gamma_c$ & Estimate & S.E. \\
\midrule
$c=1$ & $1.4423$ & $0.0203$ &
$c=1$ & $1.2547$ & $0.0891$ &
$c=1$ & $0.1323$ & $0.0145$ \\
$c=2$ & $1.6063$ & $0.0294$ &
$c=2$ & $1.4097$ & $0.1597$ &
$c=2$ & $1.1904$ & $0.2506$ \\
$c=3$ & $1.3916$ & $0.0316$ &
$c=3$ & $1.9428$ & $0.0845$ &
$c=3$ & $0.3113$ & $0.0782$ \\
$c=4$ & $1.5490$ & $0.0096$ &
$c=4$ & $1.7819$ & $0.0656$ &
$c=4$ & $0.4591$ & $0.0377$ \\
$c=5$ & $1.3470$ & $0.0348$ &
$c=5$ & $0.7340$ & $0.0661$ &
$c=5$ & $0.7067$ & $0.0951$ \\
$c=6$ & $1.5189$ & $0.0078$ &
$c=6$ & $1.1684$ & $0.0251$ &
$c=6$ & $0.1809$ & $0.0081$ \\
\bottomrule
\end{tabular}
\par\medskip
\begin{minipage}{\linewidth}
\footnotesize
\textit{Notes:} 
$^{\dagger}$Fixed by constraint: $\varphi_1=1$ and $\sigma_J=0$
(so $\log(\sigma_J)=-\infty$).
\end{minipage}
\end{table}

\end{document}